\documentclass{sig-alternate}
\usepackage{defpaper2e}

\usepackage{graphicx}
\usepackage{color}
\usepackage{amsmath,epsfig}
\usepackage{epstopdf}
\usepackage{multirow} 
\usepackage{amssymb,amsmath}
\usepackage{hhline}
\newcommand{\crowdpet}{\textit{DEPET}}

\title{Leveraging Crowd for Game-based Learning: A Case Study of Privacy Education Game Design and Evaluation by Crowdsourcing}

\numberofauthors{1} 

\author{
\alignauthor
Wendy Wang$^1$, Yu Tao$^2$, Kai Wang$^3$, Dominik Jedruszczak$^1$, Ben Knutson$^1$ \\
       \affaddr{Department of Computer Science$^1$, the College of Arts and Letters$^2$}, School of Management and Marketing$^3$\\
       \affaddr{Stevens Institute of Technology$^{1,2}$, Kean Univeristy$^3$}\\
       \affaddr{Hoboken, NJ$^{1,2}$, Union, NJ$^{3}$}\\
       \email{hwang4,ytao,djedrusz,bknutson@stevens.edu$^{1,2}$, wangkaicv@gmail.com$^3$}
}

\begin{document}

\maketitle

\begin{abstract} As the Internet grows in importance, it is vital to develop methods and techniques for educating end-users to improve their awareness of online privacy.  
Web-based education tools have been proven  effective in many domains and  have been increasingly adopted by many online professional and educational services. 
However, the design and development of Web-based education tools for online privacy is still in the early stage. The traditional solutions always involve privacy experts who have sophisticated expertise. Such  involvement can make the tool development costly. Furthermore, it is not clear how inspiring and effective these education tools are to general users of varying backgrounds, specially to novice users who have rarely dealt with online privacy issues before. In this paper, we design, develop, and evaluate a game-based privacy learning system by leveraging the wisdom of a crowd of non-experts on Amazon Mechanic Turk. Empirical
study demonstrates that the crowd can provide high-quality ideas of designing and developing a practical, educational privacy learning game. 
\end{abstract}

\section{Introduction}
\vspace{-0.05in}
Online social networking communities have undergone an explosion in recent years, as both the kinds and numbers of sites have grown and their memberships increased. This has led to the proliferation of personal data revealed by users of online communities, which presents a variety of privacy risks. 
Internet users still know little about online privacy \cite{Albrechtslund-08},  even though their awareness of online privacy has increased \cite{hargittai2010facebook}. 
Due to the fact that the existing privacy education tools are either too technical or impractical \cite{online}, there is a need for new, effective privacy education tools for Internet users of all ages and backgrounds. 

In this paper, we consider the privacy education that takes the format of game-based learning. Game-based learning is the use of digital games with educational objectives. It provides effective, interactive experiences that motivate users to actively engage in the learning process when playing with a game \cite{prensky2005computer}. However, developing an effective privacy learning game is challenging. The traditional solutions always involve privacy experts who have sophisticated expertise, deep technical skills, and broad experience. However, the involvement of professionals can make the game development costly. Furthermore, it is not clear how motivating, inspiring, and/or effective these education games are to general users of varying backgrounds, experience, and expertise of privacy, especially to novice users who have rarely dealt with privacy issues before. 

Besides practicality, creativity is another important design factor of an effective game-based learning system \cite{sisarica2013emerging}. 
Facilitated by advances in web-related technologies, crowdsourcing has become a new technique for practicing open innovation, for example, the sheep market \cite{sheep}, T-shirt design \cite{brabham2009moving}, and software idea competition \cite{leimeister2009leveraging}. It has shown that crowdsourcing workers indeed can perform some creative tasks to reduce the production costs \cite{Li-cogsci13,Sakamoto-icis11}. 

Besides the development of creative systems, crowdsourcing also has been proven effective for the evaluation of system performance. Examples of evaluation tasks include relevance evaluation for information retrieval \cite{alonso2008crowdsourcing}, book search evaluation \cite{kazai2011crowdsourcing}, and inference rule evaluation \cite{zeichner2012crowdsourcing}. Few work has used crowdsourcing to evaluate the effectiveness of a game-based learning system. 

The aim of this paper is to explore the potential of using a crowd of non-experts to help design and evaluate a privacy-education game whose target players are Internet users of all ages and backgrounds. 
There are several challenges of utilizing crowdsourcing for both design and evaluation of a game-based privacy learning system. 
First, a crowdsourcing system (e.g., Amazon Mechanical Turk \cite{turk}) typically supports microtasks that are simple, fast-paces, and require the least amount of training. It is not clear whether the crowd can perform the task such as creating a privacy education game that is complex, time-consuming, and requires some background knowledge in online privacy. 
Second, the crowd workers have various backgrounds and knowledge of online privacy. They may create game design ideas that are dramatically different in terms of both content and format. It is difficult to aggregate these ideas with significant individual heterogeneity. Third, it is challenging to evaluate the quality of the crowd's ideas, as well as the effectiveness and educational impact of a privacy education game, with  quantitative measurements. 

To address these challenges, we develop a prototype of $\crowdpet$, {\bf D}evelopment and {\bf E}valuation of {\bf P}rivacy {\bf E}ducation {\bf G}ame. The system
consists of three phases, {\em idea collection}, {\em game development}, and {\em game evaluation}. In the {\em idea collection} phase, $\crowdpet$ collects and aggregates the ideas for designing a privacy education game from a crowd of non-experts on Amazon's Mechanic Turk. In the {\em game development} phase, the privacy education game is implemented based on the aggregated ideas from the {\em idea collection} phase. In the {\em game evaluation} phase, the developed privacy game is released to both the crowd and privacy experts for effectiveness evaluation. The evaluation feedback information is then used to further improve the game.

\nop{
\begin{figure}
\centering
\includegraphics[scale=0.3] {framework.eps}
\vspace{-0.2in}
\caption{\small{\label{framework} The Framework of $\crowdpet$}}
\vspace{-0.2in}
\end{figure}
}


Our contributions include the following. First, we are the first who exploit crowd innovation for the game-based privacy learning. Second, we investigates the main characteristics of a wise group that is capable of developing creative and high-quality ideas for the development of an effective privacy education game. Third, we develop a prototype of privacy education game that instantiates the crowd's ideas. Fourth, we develop various metrics for assessing the quality of the crowd-generated ideas in terms of correctness and novelty. Fifth, we leverage both privacy experts and a large crowd of non-experts on Amazon's Mechanic Turk to evaluate the quality of the privacy education game with regard to its usability and educational impact. The evaluation results show that crowd can provide high-quality inputs for both creative system design and system evaluation. 

The paper is organized as following. Section \ref{sc:pre} defines the scope of tasks that are to be assigned to the crowd. Sections \ref{sc:collection}, \ref{sc:develop} and \ref{sc:eval} explain the details of the {\em idea collection}, {\em game development} and {\em game evaluation} phases respectively. Section \ref{sc:exp} presents the details of experiments. Section \ref{sc:related} discusses the related work. Section \ref{sc:conclusion} concludes the paper. 

\section{Definition of Problem Scope}
\label{sc:pre}

In this section, we define the scope of the privacy education game. The scope is defined at two dimensions: (1) the {\em privacy} dimension which addresses the privacy issue that the education game targets at, and (2) the {\em design space} dimension which defines the game design elements desired from the crowd. Next, we discuss the details of these two dimensions. 

\subsection{Privacy Domain}
Internet privacy is a broad term that refers to a variety of factors, techniques and technologies used to protect sensitive and private data, communications, and preferences. An education game that covers such a broad concept is expected to be complicated and cumbersome. Therefore, in this project, we narrow down to a specific privacy domain. Which privacy domain to pick is decided by two factors: (1) whether the picked domain can lead to an effective privacy education game to general users; and (2) whether the picked domain is suitable for the crowdsourcing task that requires open innovation design ideas. 
Taking both factors into consideration, we pick a specific privacy issue that has received considerable concerns by Internet users: {\em privacy issues of social
networking sites}. 
We pick this issue due to the following two reasons. 
First, a key requirement of a good game-based learning tool is to create an effective game-based learning environment that users are familiar with \cite{prensky2005computer}. According to a recent survey, 74\% of online adults use social networking sites \cite{social}.
Therefore, a game that focuses on privacy issues of social networking sites meets the requirement. It should enable general users to enjoy the game by leveraging their real-life social networking experience. 
This also should enable general users to get highly engaged in the learning process and can easily transfer what they learned from the simulated game environment to their real life. 
This addresses Factor 1. 
Second, in general, designing game-based learning tools requires the designers to be equipped with knowledge or related experience. Common crowdsourcing workers are non-experts in privacy. However, given the popularity of social networks, we expect that non-expert crowd workers can contribute high-quality design ideas of the privacy learning game  by leveraging their real-life experiences of using social networking sites. This addresses Factor 2. 


\subsection{Design Space} 

Game design is a complex process. It involves many components, including rule development, design of story plot, grading and awarding strategy, just to name a few. 
A {\em design space} identifies the range of possible solutions for a design problem \cite{brooks2010design}. The definition of design space in the crowdsourcing task is important for the quality of the collected results. A design space that is too broad and open-ended may not be able to harvest high-quality 
 ideas, while the tasks that are too closed-ended may leave too little design space for the workers to generate innovative ideas. 
Due to the fact that the types of tasks accomplished through typical crowdsourcing systems (e.g., Amazon Mechanical Turk)  have been limited to those that are low complexity, independent, and require little time and training to complete \cite{kittur2011crowdforge}, 
asking the workers to design a whole privacy education game from the scratch is too difficult to accomplish. 
Therefore, we define the game rules and GUI interface by ourselves, and ask the workers to fill in the game content only. More details of the game content will be discussed in Section \ref{sc:collection}.

\section{Idea Collection and Analysis}
\label{sc:collection}

In the {\em Collection} phase, we generated Human Intelligence Task (HIT) on Amazon Merchanic Turk (AMT), asking workers to contribute ideas of the content of a privacy education game. 
In particular, the required game content includes: 
\begin{enumerate}
\item Examples of postings shared on social networking sites (e.g., Facebook) that may expose private and sensitive
aspects of one's personal life; and 
\item Suggestions of how to fix the postings to remove the privacy information.  
\end{enumerate}
In the HIT, each worker read a preview consisting of the instructions and an example of privacy game interface. The preview was the same for all workers. The workers' answers were evaluated by the privacy experts. All workers received the payment of \$0.10 after they submitted their ideas. 

\subsection{HIT Design}

\begin{figure}[!ht]
\centering
\includegraphics[scale=0.6] {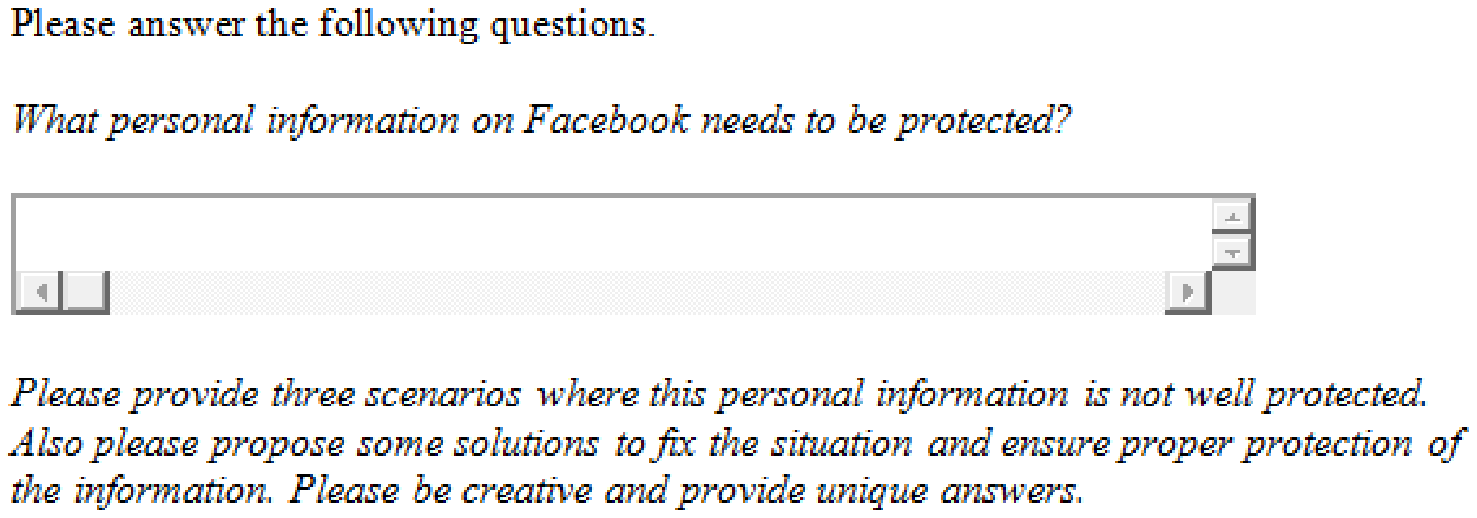}
\\
\vspace{0.1in}(a) HIT task description
\\
\includegraphics[scale=0.5] {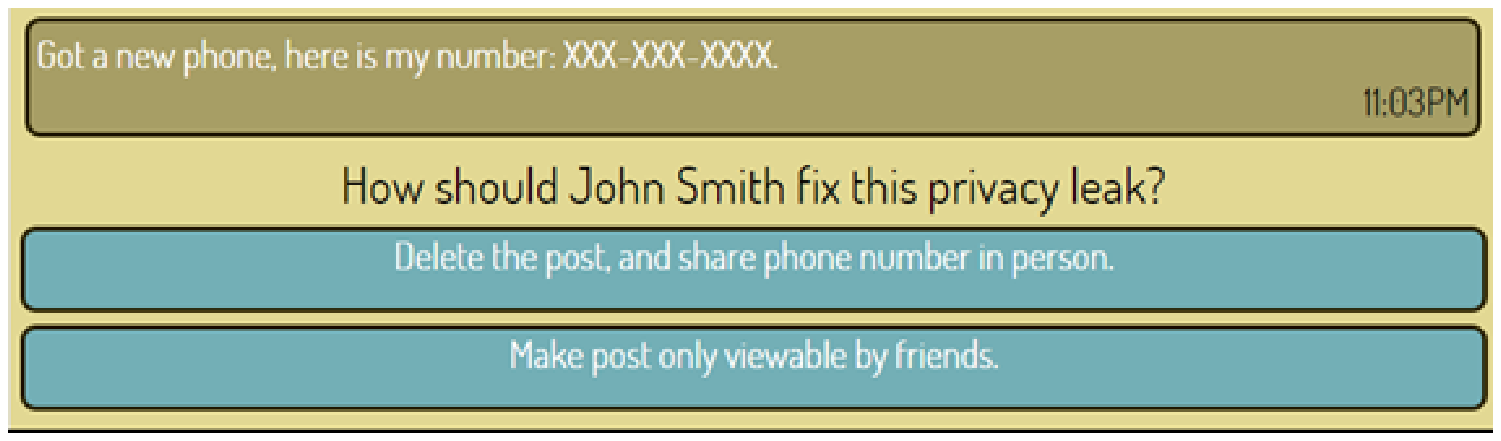}
\\
\vspace{0.1in}(b) An example of online posting used in HIT
\caption{\small{\label{example} HIT of the open-ended experiments}}
\end{figure}

Our HIT task description consists of three parts: 
\begin{itemize}
\item In {\em Preview} part, we describe the purpose of the task, and the rewarding strategy.
\item In {\em Instruction} part, we present the detailed instructions of the task. 
\item In {\em Background} part, we collect the workers' demographic information and their background knowledge in online privacy. 
\end{itemize} 

We designed two HIT tasks under different settings, namely the {\em open-ended} setting and the {\em closed-ended} setting. 

{\bf Open-ended setting.} By this task, we asked the workers to contribute any example of personal information on Facebook that needs to be protected. Each worker is asked to provide three examples of online postings that revealed some sensitive personal information. Each workers is also asked to provide fix solutions for these examples, one solution for each example. In HIT description, we emphasize {\em creativity} as one of the requirements of the examples. Figure \ref{example} (a)  shows a part of instructions that we used for the open-ended setting. To help the workers understand the task requirement and motivate them to contribute high-quality ideas, in the instructions, we include an example of online posting that reveals a person's private phone number. We also include two possible fix solutions of the problematic posting in the example. Figure \ref{example} (b) shows the example that we used in the HIT instruction. 


{\bf Closed-ended setting.} By this task, we focus on {\em four specific} privacy topics: (1) medical information, (2) income, (3) work history, and (4) student records, and asked the workers to contribute a number examples of online postings that reveal privacy information of any of these four types. We picked these four specific types of privacy information as these are the ones that the results of the {\em open-setting} experiments did not cover at all (more analysis of the {\em open-ended} experiment results can be found in Section \ref{sc:integration}). 

The task descriptions of the open- and closed-ended settings have the same {\em Preview} and {\em Background} parts; they only differ at the Instruction part. 
Due to the disjoint types of privacy information that were covered by both open- and closed-ended experiments, we allow the same workers to participate in both experiments. Both open- and closed-ended settings have the same rewarding mechanism. 

\noindent{\bf Workers' background.} As a part of HIT description, we collected the workers' demographic information,  including their locations, gender, and education level. We also surveyed their background knowledge in online privacy, including: (1) how often do they use Facebook; (2) whether they have adequately protected their own private information on Facebook; and (3) how much do they know about online privacy. We collect these information trying to find out the correlation between the quality of ideas and the workers' background. 

\subsection{Idea Aggregation and Analysis}
\label{sc:integration}

\vspace{-0.05in}
\subsubsection{Idea Aggregation}

Sixteen workers participated in the idea generation phase, each contributing three example postings that revealed some sensitive personal information. We observe that these forty-eight examples share a common set of privacy topics. 
Therefore, we design an idea aggregation method that enables to automatically aggregate the collected crowd-generated design ideas. A possible solution is to use topic modeling \cite{papadimitriou1998latent}. However, there are several challenges of using topic modeling to aggregate crowd's ideas. First, many existing topic modeling methods (e.g., LDA \cite{blei2003latent}) rely on a large document corpus to get decent results, while the idea description collected from AMT are short (no more than 200 characters in our case) and abbreviated. The number of ideas is also small (48 examples), given our limited budget to hire AMT workers. Second, the crowd's input is noisy - it contains typing errors and meaningless words. Due to these two challenges, applying the existing topic modeling methods to the crowd's ideas fails to deliver good results. Therefore, instead of using any topic modeling method based on training and learning, we use the keyword extraction method based on word frequency analysis. We use TF-IDF to weight the terms. Then we identify those terms that are appropriate as privacy topics. We found thirteen distinct privacy topics by this method. 
We grouped the examples provided by the crowd according to the privacy topics that they are associated with.
It turned out that the privacy topics {\em activity}, {\em home address}, and {\em personal emotions} are the three that are associated with the highest number of examples, while the popular privacy topics such as {\em name}, {\em age}, and {\em birth date} has fewer examples. This is surprising as the privacy concerns of revealing activities and personal emotions on social networking sites are indeed paid much less attention as name and birth date \cite{huangprivacy}.
This shows that the workers indeed intended to contribute creative ideas of privacy examples, as required in the HIT instructions. 

\vspace{-0.05in}
\subsubsection{Idea Analysis}

In this subsection, we analyze the aggregated ideas for both open-ended and closed-ended settings. 

\noindent{\bf Open-ended Setting.}  Sixteen workers participated in this round of experiments, each contributing three example postings that revealed some sensitive personal information. There are forty-eight examples in total. We analyze these examples and have the following interesting observations. 

\begin{figure}[!ht]
  \begin{center}
	  \includegraphics[width=0.4\textwidth]{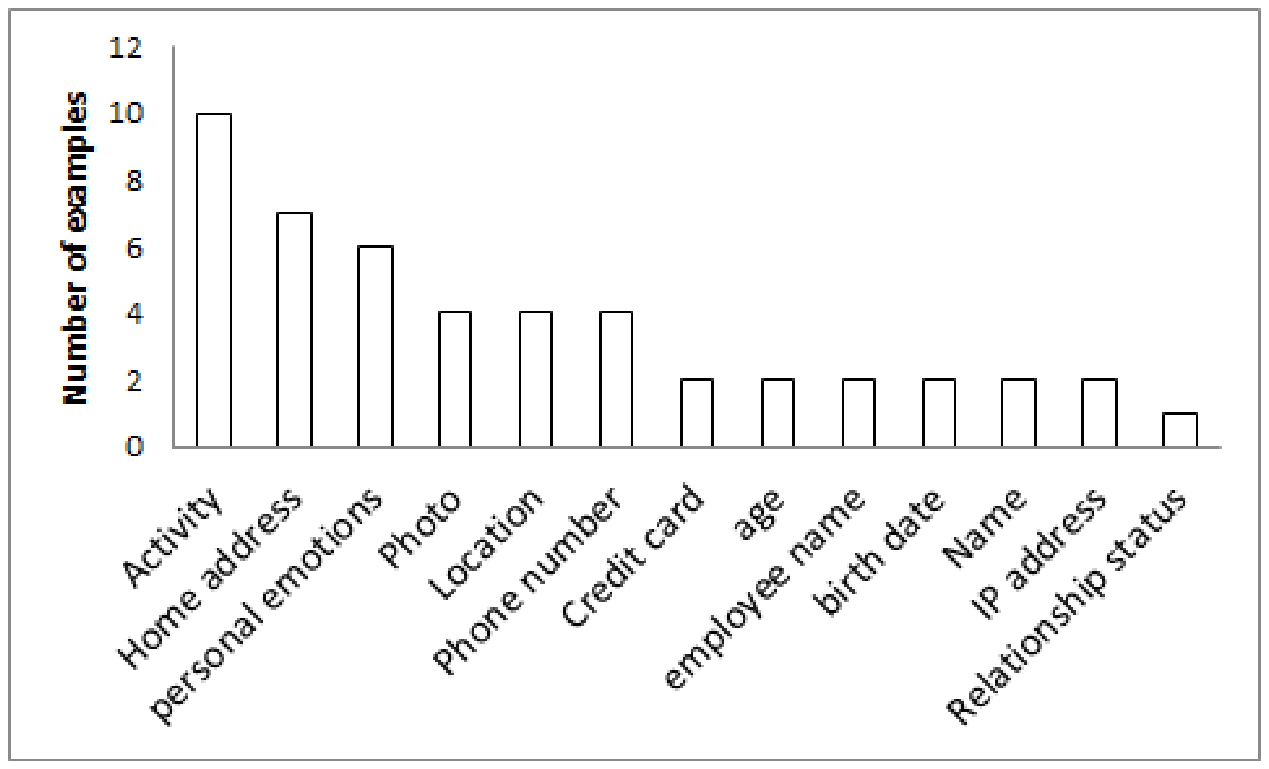}
	\\
		(a) Number of examples of each privacy topic
  \\		
    \includegraphics[width=0.4\textwidth]{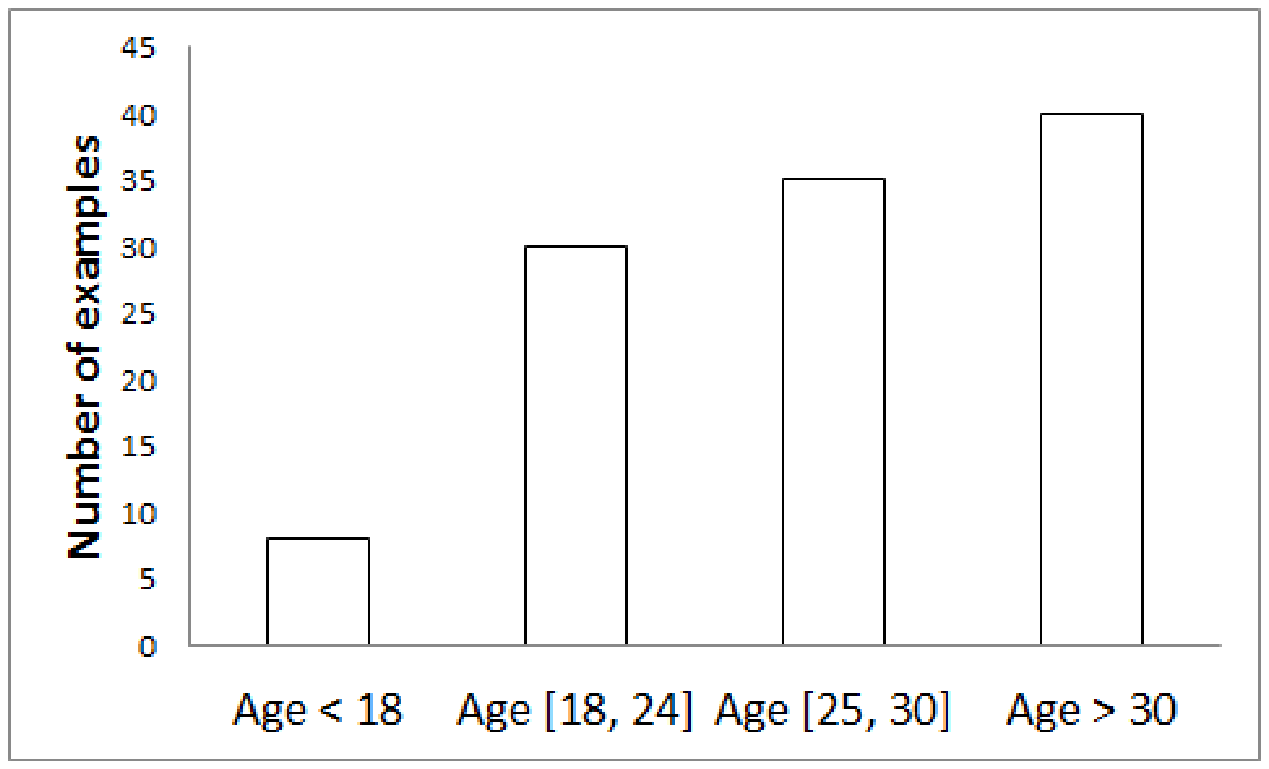}
	\\
		(b) Number of examples for each population group
    \caption{\small \label{fig:topic} Analysis of Crowd-generated Ideas}
  \end{center}
\end{figure}

First, there exists a large overlap on the type of {\em privacy topics} that the examples covered. We categorized the examples by their privacy topics, and counted the number of examples that each privacy topic has. There are thirteen distinct privacy topics in total. Figure \ref{fig:topic} (a) shows these privacy topics and the number of examples that each privacy topic covers. Among these topics, {\em activity}, {\em home address}, and {\em personal emotions} are the three topics that are of the highest frequency, while the popular privacy topics such as name, age, and birth date has fewer examples. This is surprising as the privacy concerns of revealing activities and personal emotions on social networking sites are indeed paid much less attention as name and birth date \cite{huangprivacy}.
This shows that the workers indeed intended to contribute creative ideas of privacy examples, as required in the HIT instructions. 

\begin{figure*}[ht]
  \begin{center}
	  \begin{tabular}{cc}
	  \hspace{-0.3in}\includegraphics[height=1.4in]{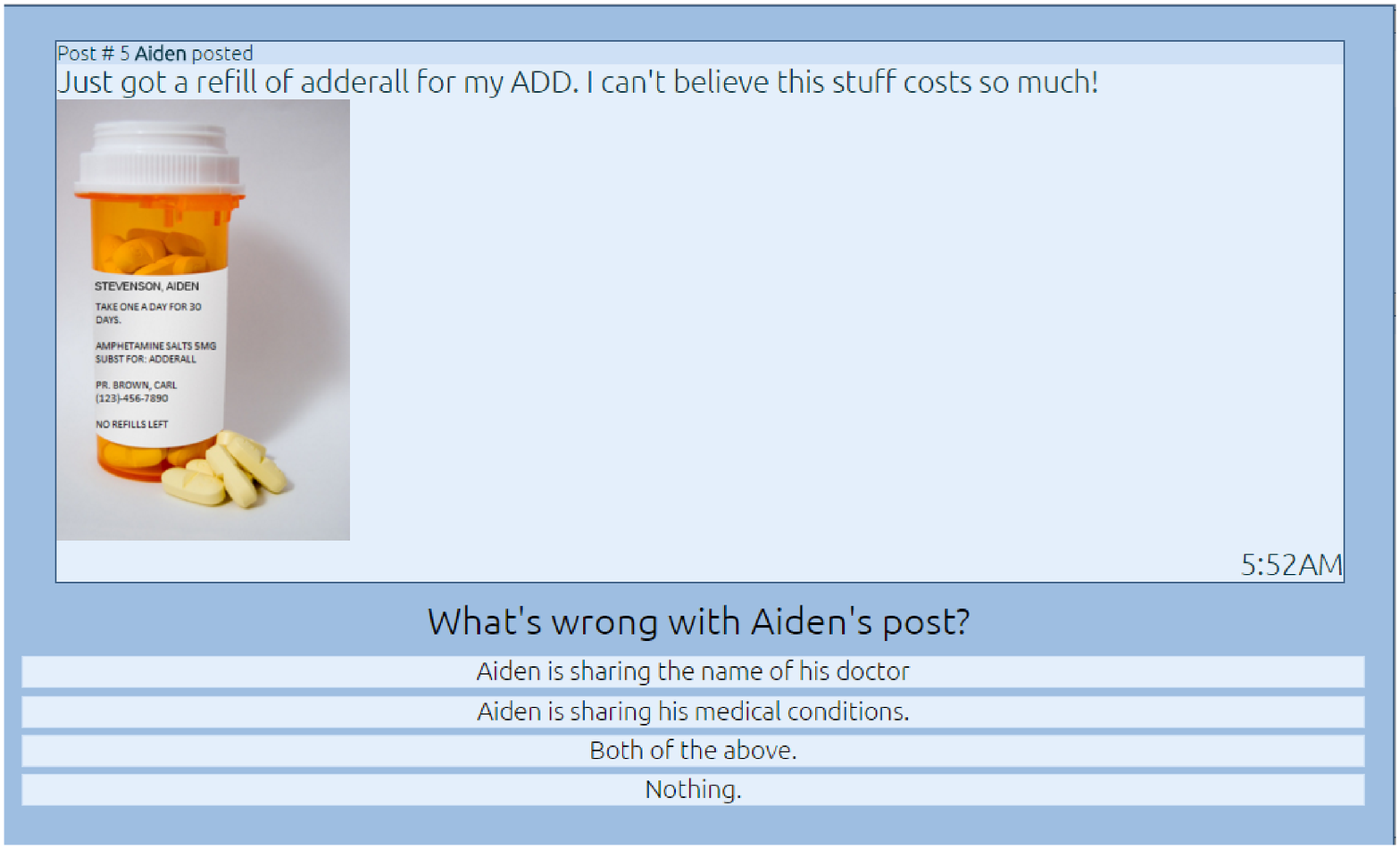}
		&
    \hspace{-0.4in}\includegraphics[height=1.4in]{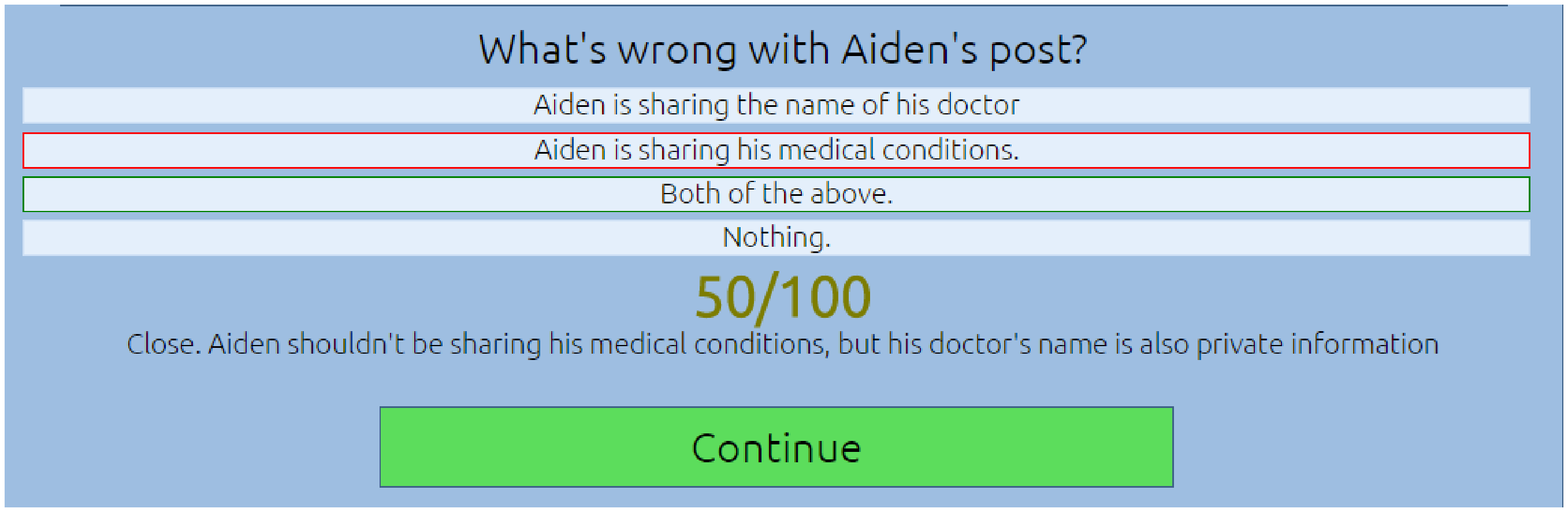}
		\\
		(a) An example of posting that has privacy issues
    &		
		(b) An example of the feedback of privacy issues
		\end{tabular}
		\\
		\begin{tabular}{cc}
		\includegraphics[height=1.4in]{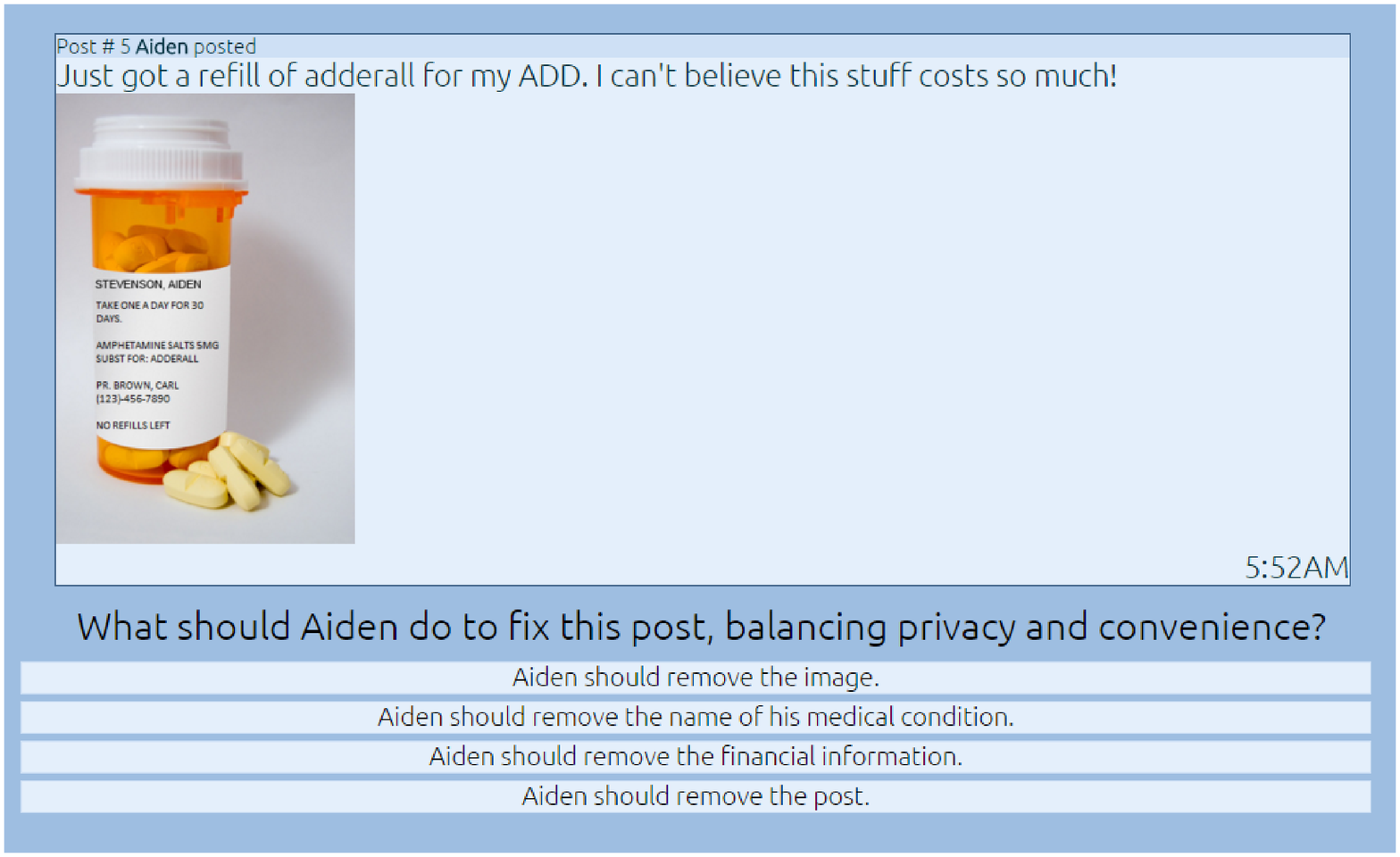}
		&
		\includegraphics[height=1.4in]{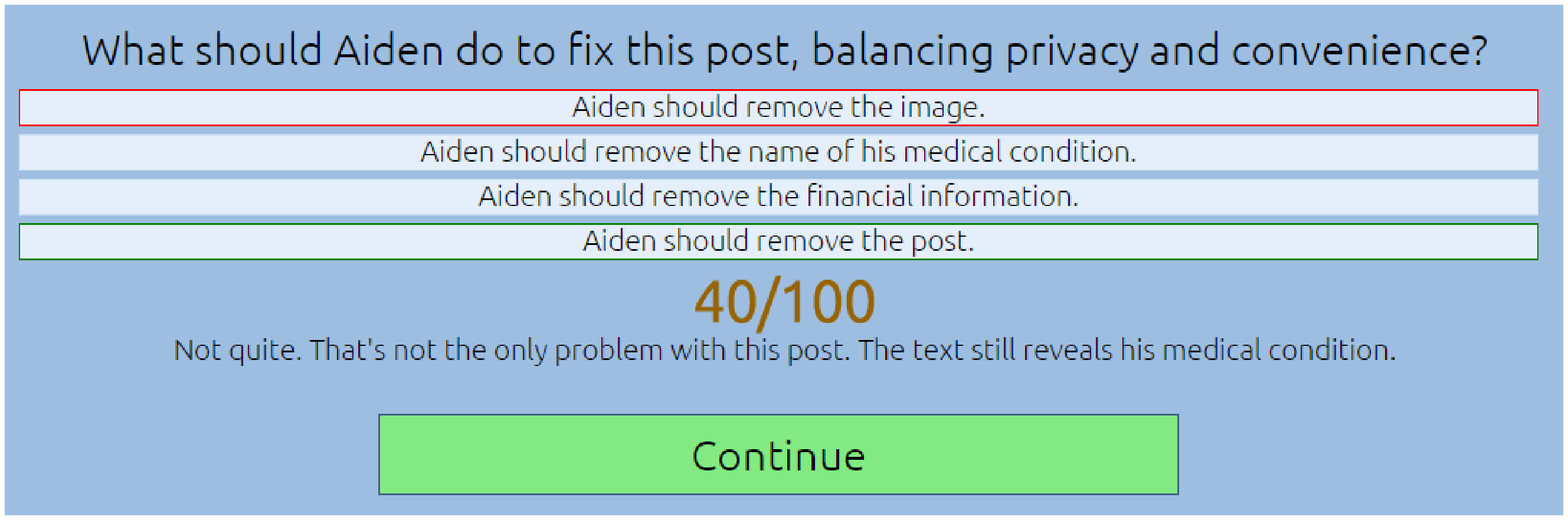}
		\\
		(c) An example of fix solutions
		&
		(d) An example of the feedback of picked solution
		\end{tabular}
    \caption{\small \label{fig:game} Example Posting and Possible Fix in Game}
  \end{center}
\end{figure*}
Second, the crowd-generated example postings are very diverse in terms of the population groups that the examples can apply to. 
The examples that belong to the same privacy topic group indeed target on different group populations. For instance, in group of the privacy topic {\em location}, one example described an adult's posting showing his/her exercise routine at a specific gym, while another example mentioned the location of a kid's favorite  playground. We categorize the workers' examples into four population groups: (1) kids and high school students (age $<18$); (2) college students (age in [18, 24]); (3) young professionals (age in [25, 30]); and (4) older adults (age $>30$). We assign the examples to these four population groups, and count the number of examples in each group. The grouping of examples is not exclusive; some  examples can be applied to multiple groups. Figure \ref{fig:topic} (b) shows the grouping results. It turned out that the {\em older adult} group received the highest number of privacy topics. This may due to the fact that a majority of workers are in the same age range as the older adult group; they design the examples from their real-life experiences. 

\noindent{\bf Closed-ended Setting.} 
In this round, twelve workers participated, each creating three example postings. There are thirty-six examples in total. These examples cover four specific types of online privacy topics (as required in HIT), namely medical information, income, work history, and  student records. Each topic has nine examples. We found out that none of the examples are similar. More details of the similarity measurement and the analysis of the example similarity can be found in Section \ref{sc:quality}.  Therefore, we aggregate the ideas collected from the closed-ended setting based on the four privacy topics.

\vspace{-0.1in}
\section{Game Development}
\label{sc:develop}

In this phase, we implemented a prototype of our Web-based privacy education game $DEPET$\footnote{The prototype is available at http://stevens.edu/depet.}. The ideas that were collected and integrated from the crowd were implemented as Facebook-style posting examples in the game. $DEPET$ was implemented by using MySQL, Javascript, and HTML5. 

According to the four different target groups of population that the crowd-generated example postings covered, we designed four different characters in the game. 
Each character is associated with a set of privacy topics. The privacy topics that each character is associated with are:  
\begin{itemize}
\item Sophia (high school student, age$<18$): ID, credit cards, driver license, and academic behaviors;
\item Aiden (college student, age $[18, 24]$): home address, academic behaviors, medical information, and personal emotions;
\item Olivia (young adult, age $[25, 30]$): professional issues; 
\item Lucas (older adult, age $>30$): information of family members, personal ideology, and medical information. 
\end{itemize}


We design the following character-based game rules. The players can pick any of the four aforementioned characters to start the game. Each game character has five or six online postings, each posting revealing some private and sensitive
aspects of one's personal life. Each posting is also associated with three possible fix solutions. 
Most of the example postings and the fix solutions in the game are generated from the {\em idean collection} phase. The players have to: (1) decide whether these postings have privacy issues, and what these privacy issues are if there is any; and (2) what is the best fix solution in terms of balanced privacy and data availability. 
Users' different decision of where the privacy problem occurs is assigned with a different performance score. 
An example of the postings is shown in Figure \ref{fig:game} (a). 
Besides showing the scores, the game gives the players the feedback of their choices. The feedback advises the players why their choices are (in)correct, and what are the real privacy problems of the postings. An example of the feedback is shown in Figure \ref{fig:game} (b). The feedback was designed by the privacy experts.  
After the players finishing judging the privacy problems of the postings, the game shows three to four fix solutions.  
An example of the fix solutions is shown in Figure \ref{fig:game} (c).  
The best solution is the one that best addresses the trade-off between privacy and data availability. 
The players pick one solution. Each solution has a score, depending on how good privacy and data availability is balanced. For instance, consider the four solutions shown in Figure \ref{fig:game} (c). Removing the image alone (Option A) or only the specific medical condition in the text (Option B) cannot protect the privacy entirely. Therefore, the only correct solution is to remove the post. The game also gives the feedback to the players of their picked fix solutions. The feedback states clearly why the picked solution can(not) remove the privacy leak sufficiently.

\section{Evaluation by Crowd} 
\label{sc:eval}

In this project, we evaluated the $DEPET$ prototype in terms of its usability and education success by leveraging the crowd on Amazon Mechanical Turk. 

\subsection{Task Description}

Our task description consists of three parts: 
\begin{itemize}
\item In {\em Preview} part, we described the purpose of the evaluation task and the rewarding strategy. Each participant is paid \$0.20 after they finish the evaluation. 
\item In {\em Instruction} part, we present the detailed instructions of the task. The Web link of $DEPET$ is attached in the instructions.
\item In {\em Background} part, we collected the evaluators' demographic information,  including their locations, gender, and education information. We also surveyed their background knowledge in online privacy, including how often do they use Facebook and their self-evaluation of privacy levels. 
\end{itemize} 

\begin{figure}[!ht]
  \begin{center}
	  \includegraphics[width=0.45\textwidth]{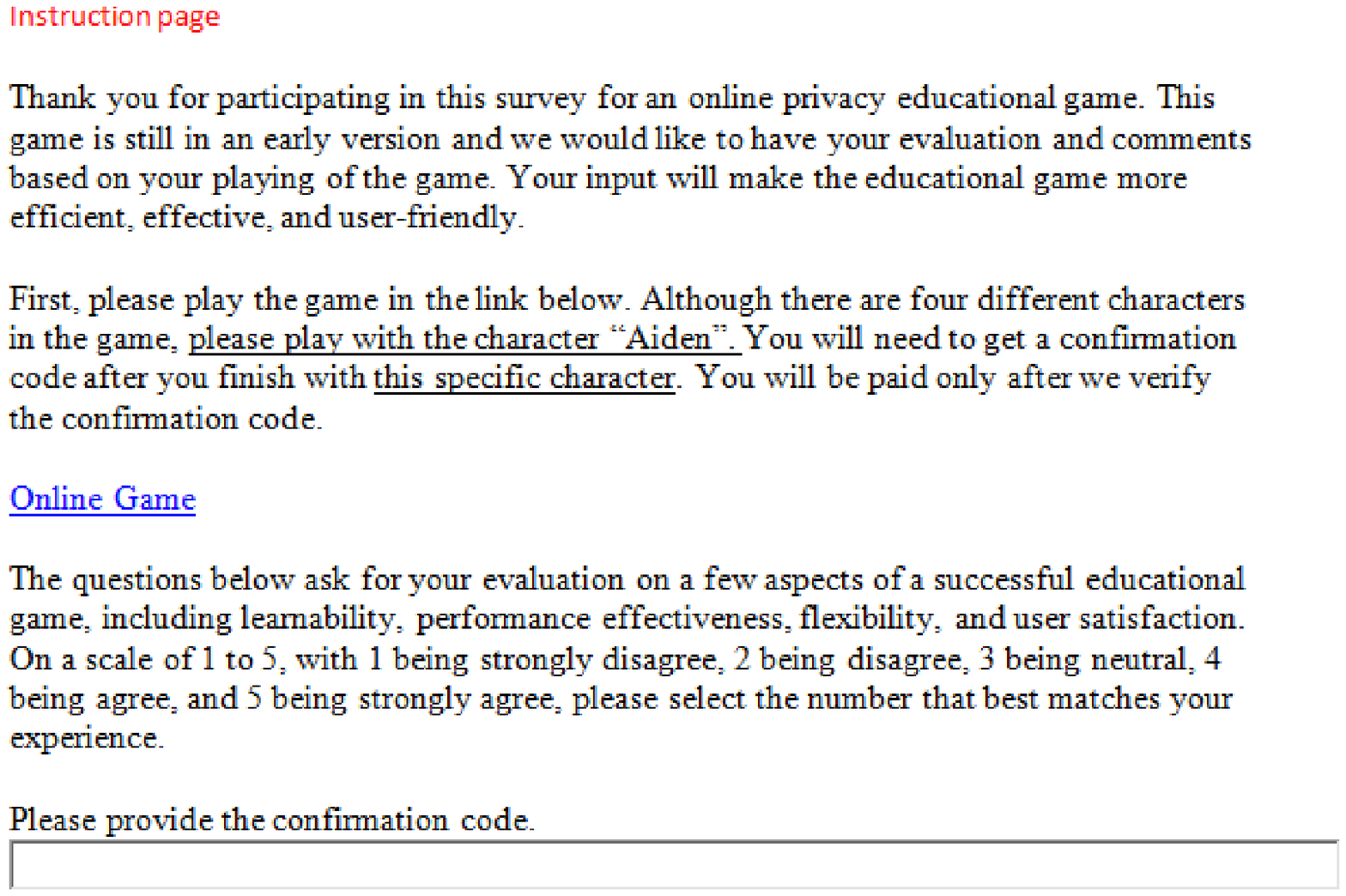}
		\caption{\small \label{fig:evalGUI} Evaluation HIT task description}
  \end{center}
\end{figure}

To ensure that all four game characters are evaluated by the same number of workers, we released four evaluation HITS, one for each character. In each HIT, we asked the player to evaluate the character as required in the task description. We require that each participant reviews and plays with all example postings and the suggested fix solutions (of one specific character) before (s)he evaluates the game. To ensure this, we created a confirmation code for each character. Only those workers who finished all example postings of the character will receive the confirmation code. Different characters have different confirmation codes. The workers are asked to enter the confirmation code before they start the evaluation. Figure \ref{fig:evalGUI} shows the part of instructions with the confirmation code field. 

The evaluation took the format of user surveys. After the workers enter the confirmation code,  they answer a set of questions in the survey regarding the game character that they played with. 
The survey evaluates two types of effectiveness of the developed game prototype for privacy education and learning: 
\begin{itemize}
\item {\em Tool usability}, which measures whether the tool (i.e., the game) is easy to use; and 
\item {\em Educational success}, which measures the impacts of the tool on delivery and learning of online privacy knowledge. 
\end{itemize}
There are twenty questions in the survey. Among these questions, fifteen questions are single-choice questions, and five questions are free-text format. The single-choice questions ask the users to select the number that best matches their experience. The numbers specifies a scale of 1 to 5, with 1 being strongly disagree, 2 being disagree, 3 being neutral, 4 being agree, and 5 being strongly agree. The free text-format questions ask the workers to enter they feedback and suggestions. 
Next, we discuss how we designed the questions in the survey. 
  
\noindent{\bf Evaluation of tool usability.} We used the {\em usability testing} method to evaluate tool usability. Usability tests, with a user-centered approach, have been successfully used in the development of other products such as interactive multimedia software and web-based learning tools \cite{sung99,shiratuddin02,pass04}. The usability test model by \cite{sung99} suggests to test: (1) Learnability (the ease of learning to use the system, e.g., clearly labeled components, easy navigation); (2) Performance effectiveness (the ease of using the system in terms of speed and error rate); (3) Flexibility (the level of freedom to use multiple commands to achieve the same goal); (4) Error tolerance and system integrity (the ability to recover from errors and prevent data corruption or loss); and (5) User satisfaction (users' opinions and perception of the training tool). 
For each testing component, we designed corresponding questions in the HIT, asking for the crowd's feedback. 

\noindent{\bf Evaluation of educational success.} Kirkpatrick proposed a training evaluation model \cite{kirkpatrick-1979techniques} that can objectively analyze the effectiveness and impact of education. We apply Kirkpatrick's evaluation model to our evaluation phase. Kirkpatrick's evaluation model measures the following four items.

{\bf Level-1 Reaction Measurement.} In the survey, we ask the questions related to what the workers thought and felt about the game. For example, did they feel that the game was worth their time? What were the biggest strengths and weaknesses of the game? 
Did they like the topic, the materials, and the format?
By measuring reaction, we will understand how well the game-based learning was received by the workers. 

{\bf Level-2 Learning Measurement.} 
To measure how much the workers' knowledge of online privacy has increased as a result of the education game, we asked the workers in the survey whether their knowledge of online privacy has increased as a result of the playing the game. 

{\bf Level-3 Behavior Measurement.} 
We ask a series of questions regarding how the workers  would apply the knowledge learned from the game. For example, will the workers put any of their learning of online privacy techniques to their practical use?
Are they able to teach their new knowledge and skills to other people? The main focus is the behavioral changes due to the game playing.

{\bf Level-4 Results Measurement.} 
We analyze the effect of the game on the improved performance of the workers. Kirkpatrick suggested several key performance indicators, such as volumes, percentages, and other quantifiable aspects of performance, such as whether the number of privacy accidents has been reduced \cite{kirkpatrick-1979techniques}. It is hard to collect such observable performance indicators in the survey. Therefore we did not ask such questions. Instead we asked the workers whether they would recommend this game to others for the learning on online privacy. 

There are twenty questions in the survey. We categorize all the questions into three components: 
\begin{itemize}
\item Evaluation of game interface, including seven questions that cover the evaluation of tool usability; 
\item Evaluation of game content, including seven questions that cover the evaluation of both tool usability and education success; and
\item Overall evaluation, including six questions that cover the evaluation of education success. 
\end{itemize}

\section{Experiments}
\label{sc:exp}

In this section, we explain the details of our experiments.

\subsection{Setting}

We performed both the idea collection and game evaluation tasks on Amazon Mechanical Turk. Besides the crowd, we also have three privacy experts: a professor whose expertise in data privacy and two of her senior PhD students.  

\subsection{Workers' Background}
\nop{
  \vspace{-0.05in}
\begin{figure}[!ht]
  \begin{center}
    \begin{tabular}{cc}
      \includegraphics[width=0.23\textwidth]{education.eps}
      &
      \includegraphics[width=0.23\textwidth]{gender.eps}
      \\
      (a) Education
      &
      (b) Gender
    \end{tabular}
    \caption{\small \label{fig:open-demo} Demographic information of workers in the open-ended setting}
  \end{center}
\end{figure}
}

\begin{figure*}[ht]
  \begin{center}
    \begin{tabular}{ccc}
      \includegraphics[width=0.23\textwidth]{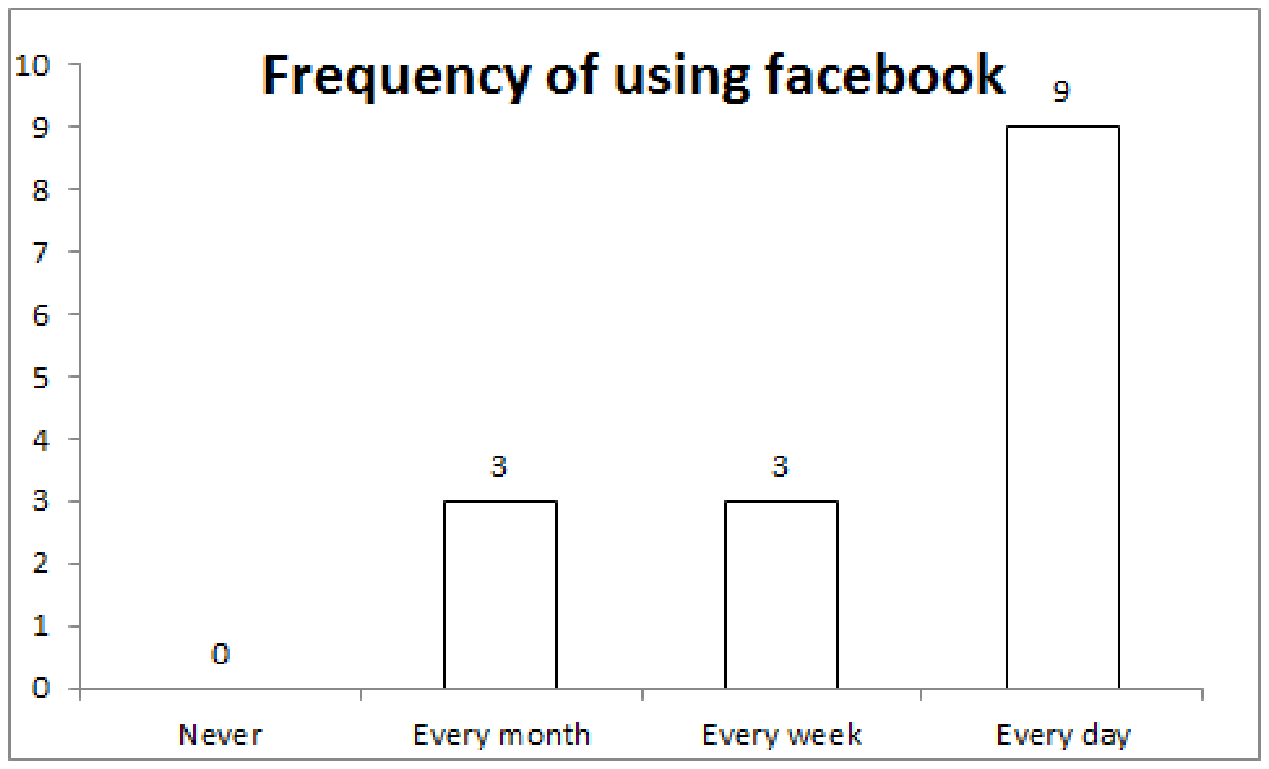}
      &
      \includegraphics[width=0.23\textwidth]{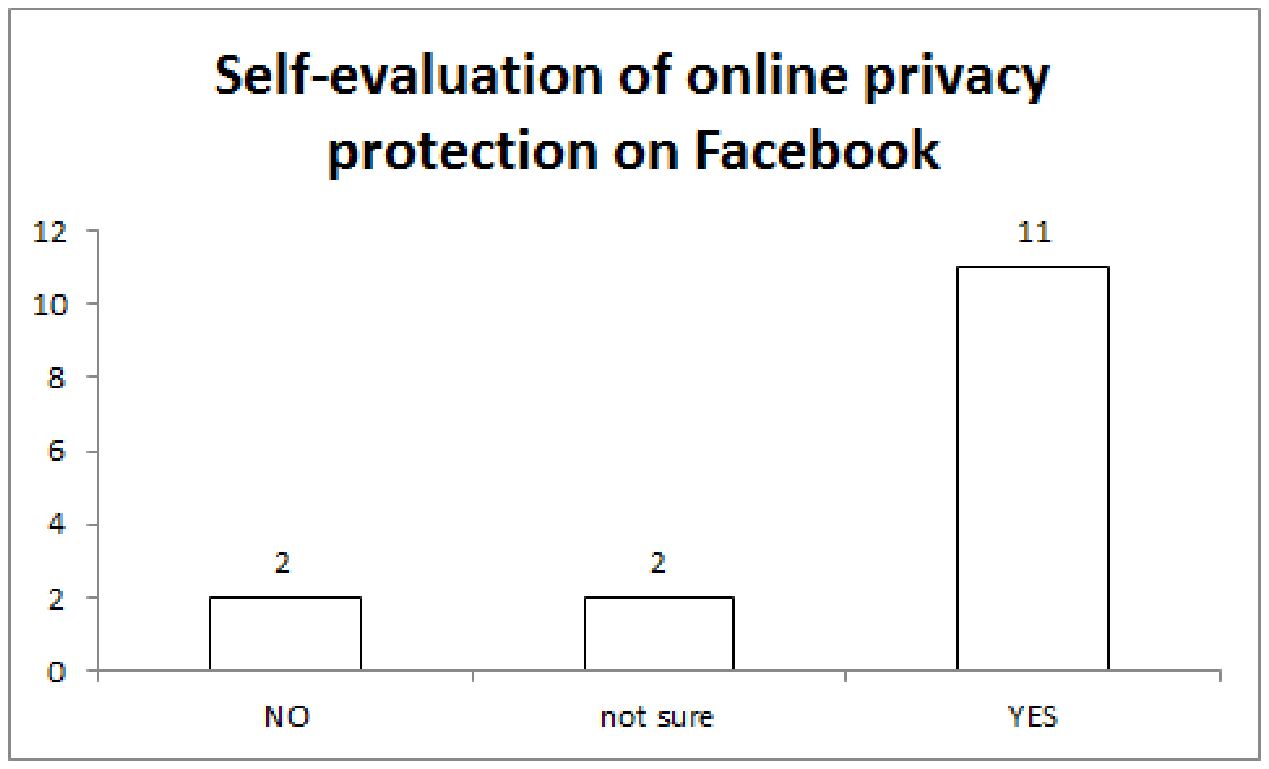}
			&
      \includegraphics[width=0.23\textwidth]{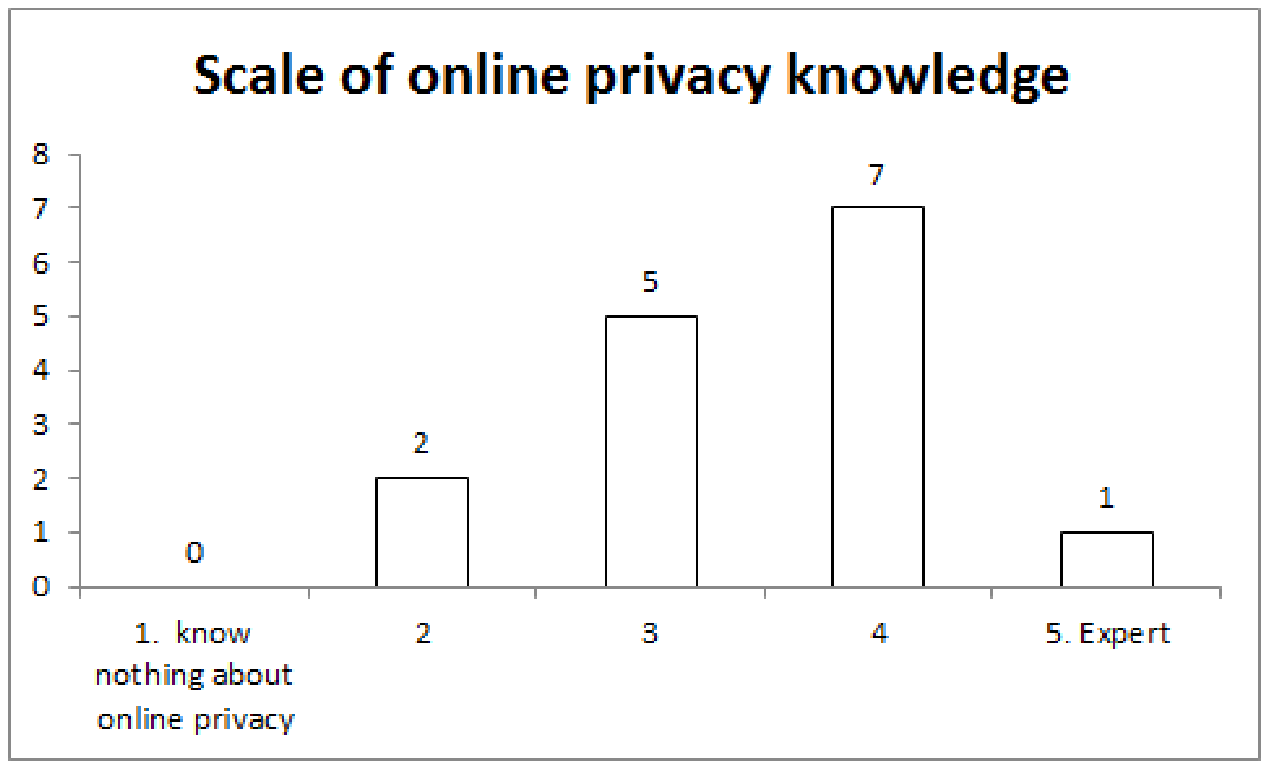}
      \\
      (a) How often use Facebook
      &
      (b) Self-evaluation of adequately protecting 
			&
			(c) Self-evaluation of privacy expertise scale
			\\
			&
			private information on Facebook
			&
    \end{tabular}
    \caption{\small \label{fig:open-pri} Privacy background and experiences of the workers in open-ended setting}
  \end{center}
\end{figure*}

\nop{
\begin{figure}[!ht]
  \begin{center}
    \begin{tabular}{cc}
      \includegraphics[width=0.23\textwidth]{set2-edu.eps}
      &
      \includegraphics[width=0.23\textwidth]{set2-gender.eps}
      \\
      (a) Education
      &
      (b) Gender
    \end{tabular}
    \caption{\small \label{fig:close-demo} Demographic information of workers in the close-ended setting}
  \end{center}
\end{figure}
}

\begin{figure*}[ht]
  \begin{center}
    \begin{tabular}{ccc}
      \includegraphics[width=0.23\textwidth]{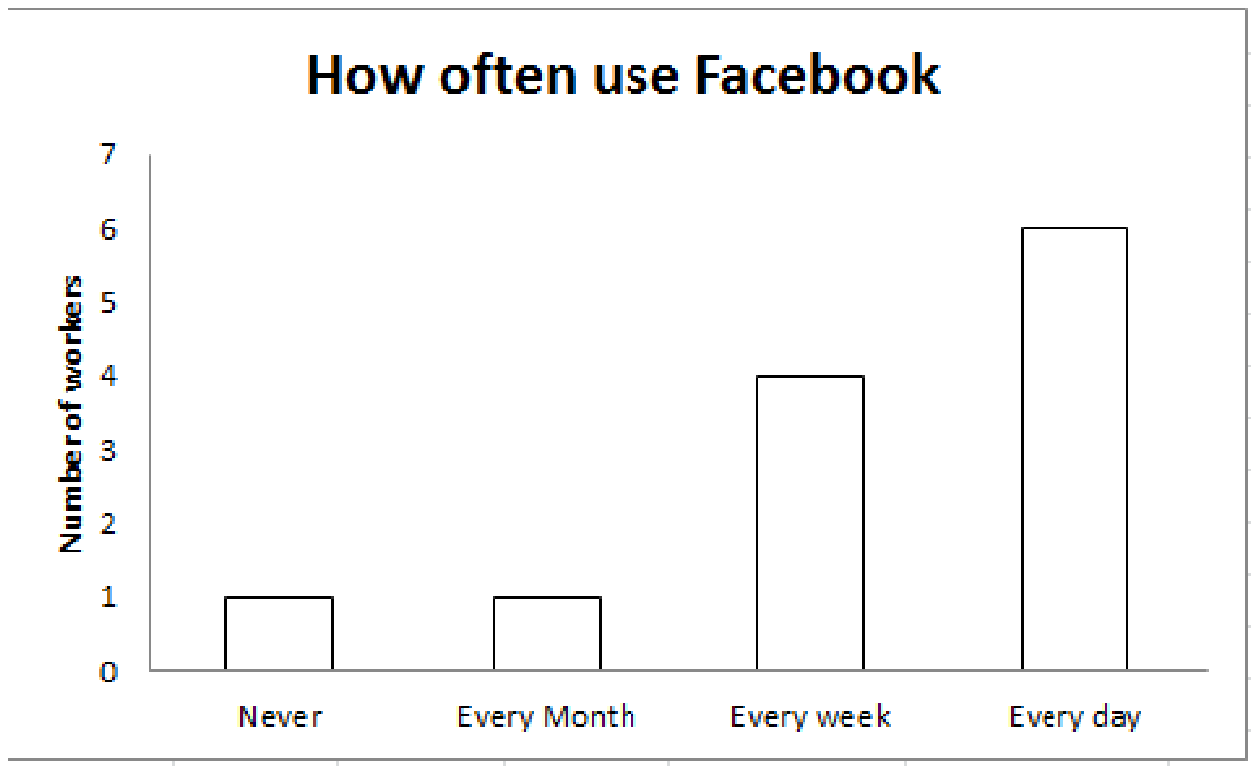}
      &
      \includegraphics[width=0.23\textwidth]{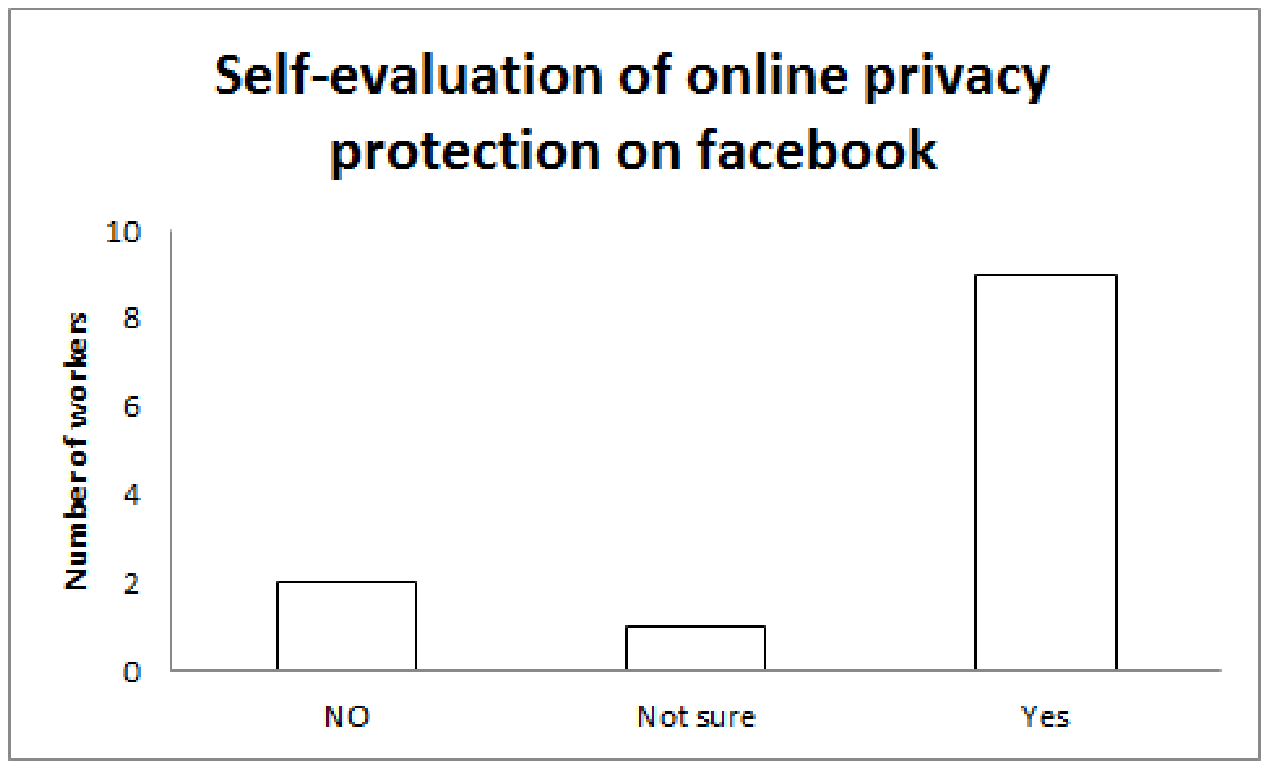}
			&
      \includegraphics[width=0.23\textwidth]{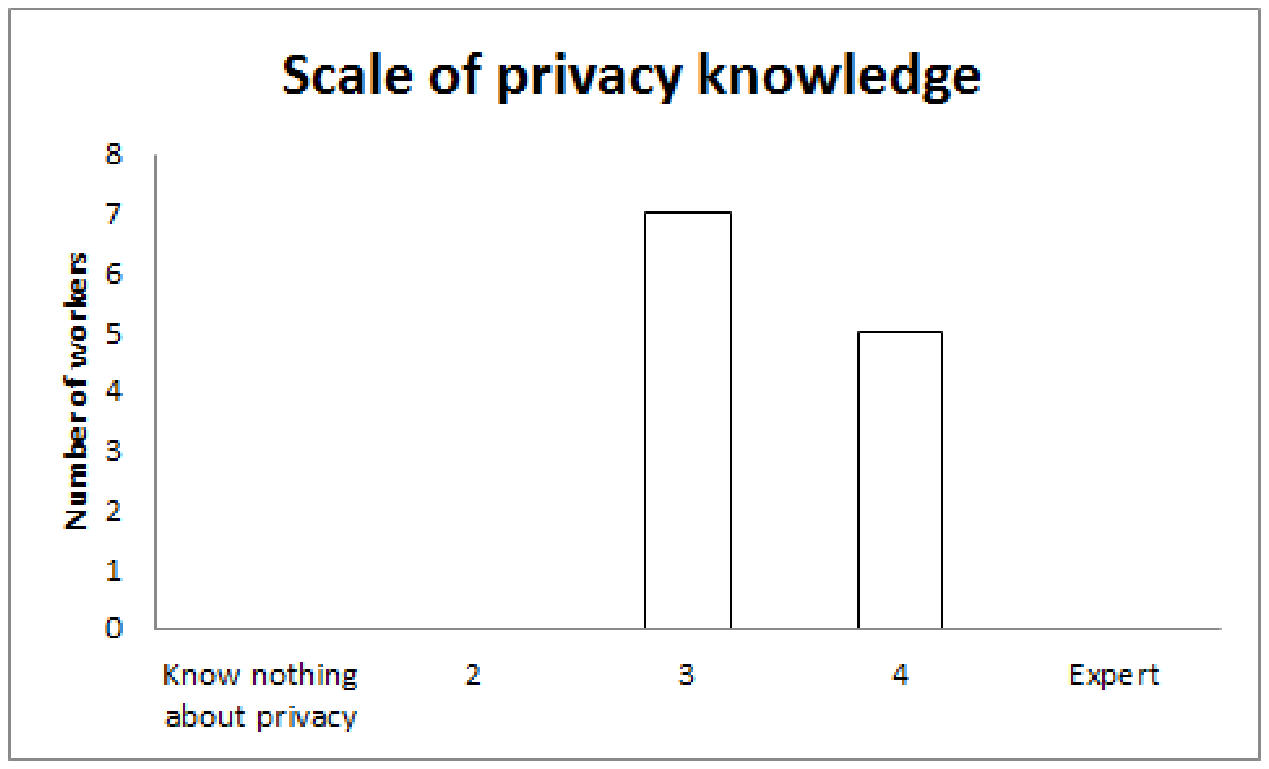}
      \\
      (a) How often use Facebook
      &
      (b) Self-evaluation of adequately protecting 
			&
			(c) Self-evaluation of privacy expertise scale
			\\&
			private information on Facebook
			&
    \end{tabular}
    \caption{\small \label{fig:close-pri} Privacy background and experiences of the workers in close-ended setting}
  \end{center}
\end{figure*}

We collected the demographic information of workers, as well as their knowledge of online privacy, for both Collection and Evaluation phases. 

\subsubsection{Idea Collection Phase}
\noindent{\bf Open-ended setting.} Sixteen workers participated this experiment. All users are from US. The mean of their ages is 39. Regarding the education level, ten workers did not finish high school. Four finished high schoo, and two had college education. Regarding the gender, ten workers are females and six were males. Regarding the background knowledge and experiences on social networking privacy, all workers use Facebook at least every month (Figure \ref{fig:open-pri} (a)). Most of them considered themselves having protected their privacy on Facebook sufficiently (Figure \ref{fig:open-pri} (b)). A majority of these workers considered themselves have intermediate level of privacy knowledge (Figure \ref{fig:open-pri} (c)). 

\noindent{\bf Closed-ended setting.} Twelve workers participated this experiment. All users from US. The mean of their ages is 26. Regarding the education level, nine workers finished high school, and three received college education. Regarding the gender, seven workers are females and five were males. Regarding the background knowledge and experiences on social networking privacy, all workers except one use Facebook at least every month (Figure \ref{fig:close-pri} (a)). Most of them considered themselves having protected their privacy on Facebook adequately (Figure \ref{fig:close-pri} (b)). All these workers considered themselves have intermediate level (scale 3/4 out of 5) of privacy knowledge (Figure \ref{fig:close-pri} (c)). 

To summarize, in both settings, the crowd who generated game design ideas are of various demographic backgrounds. They also have different knowledge and experiences in online privacy.

\subsubsection{Game Evaluation Phase}
One hundred and twenty workers participated into the evaluation task. All of them are from US. The average age is 33. Regarding the gender, sixty-two workers are females, and fifty-eight are males. Regarding the education level, 34 workers did not finish high school, 74 workers finished high school, and 12 workers received college education. Regarding their online privacy experience, 11 workers considered themselves did not protect privacy sufficiently, 85 workers considered them with sufficient protection of online privacy, and 24 workers were not sure. Regarding their privacy expertise, one worker knows nothing about online privacy, 5, 45, and 69 workers labeled themselves as level 2, 3, and 4 (out of 5) respectively. None of the workers considered themselves as an expert on online privacy. The average of self-evaluation expertise level is 3.69 (out of 5). Again, the crowd who evaluated the game came from various demographic backgrounds, with different knowledge and experiences in online privacy. 

\subsection{Metrics of Idea Quality Evaluation}
We define two metrics, namely the {\em correctness} and {\em novelty}, to evaluate the quality of the crowd-generated ideas. 

\noindent{\bf Correctness.} Informally, the correctness measurement of workers' ideas is to verify whether the ideas indeed expose private and sensitive
aspects of one's personal life (explicitly or implicitly). We refer to the privacy experts' evaluation for correctness measurement. Each idea receives a binary value (0/1) indicating its correctness. Given $m$ ideas, the correctness ratio is defined as 
\[cr = \frac{k}{m},\]
where $k$ is the number of correct ideas. Intuitively, the correctness ratio measures the portion of the ideas that are correct. 

\noindent{\bf Novelty.} The novelty measurement is based on the {\em similarity} of ideas. To measure the pairwise similarity of ideas, first, we manually extract the keywords of each idea. We only kept the meaningful keywords with stems removed. Synonyms are merged to one representative word (e.g., ``photo'' and ``picture'' are represented by ``photo''). 
Then for every two ideas $E_1$ and $E_2$, let $K_1$ and $K_2$ be their keywords, we measure the similarity of $K_1$ and $K_2$ by measuring the similarity of $E_1$ and $E_2$. There are a number of approximate string similarity measurement methods in the literature, e.g., $q$-grams, edit distance, and Euclidean distance (See \cite{koudas2006record} for a good tutorial.) 
In this paper, we consider the {\em Jaccard} similarity based on $q$-grams. In particular,
each keyword is decomposed into a set of tokens of length $q$ (called {\em $q$-grams}).
Then for any two strings $S_1$ and $S_2$,
\begin{eqnarray}
jaccard(S_1, S_2) = \frac{|G(S_1,q) \cap G(S_2,q)|}{|G(S_1,q)\cup G(S_2,q)|},
\end{eqnarray}
where $G(S_1,q)$ ($G(S_2,q)$, resp.) is the set of q-grams of $S_1$ ($S_2$, resp.).
We say two keywords  $S_1$ and $S_2$ are {\em $\delta$-similar} regarding the Jaccard metrics, if $jaccard(S_1, S_2) \geq\delta$. Given two sets of keywords $K_1 =\{S_1^1, \dots, S_m^1\}$ and $K_2=\{S_1^2, \dots, S_n^2\}$, let $\ell$ be the number of $\delta$-similar keyword pairs of $K_1$ and $K_2$, the similarity of $K_1$ and $K_2$ is measured as 
\begin{eqnarray}
\label{eqn:sim}
sim(K_1, K_2) = \frac{\ell}{|K_1\cup K_2|}.
\end{eqnarray}
We say two ideas $D_1$ and $D_2$ are similar, denoted as $D_1 \approx D_2$,  if the similarity of their keywords $K_1$ and $K_2$ satisfies $sim(K_1, K_2)\geq\theta$, where $\theta\in[0,1]$ is a user-specified threshold. 

Given a set of ideas $D_1, \dots, D_n$, the novelty of $D_i (1\leq i\leq n)$ is measured as:
\[novelty(D_i) =\frac{n - |\{D_j|D_j \approx D_i, i\neq j\}|}{n}.\]
That is, the novelty of $D_i$ measures the percentage of ideas that are dissimilar to $D_i$. Intuitively, the more the dissimilar ideas there are, the higher the novelty of $D_i$ is. 

\subsection{Idea Quality}
\label{sc:quality}
In this set of experiments, we measure: 
(1) both the correctness and novelty of the ideas collected by the Collection phase; and 
(2) the correlation between the quality of workers' ideas and their privacy background knowledge and experiences. 

\noindent{\bf  Correctness \& Novelty.} First, we measure the correctness of the collected forty-eight posting examples and their suggested fix solutions in the open-ended setting. Seven examples were evaluated as incorrect. All suggested fix solutions (of the correct examples) were correct. The correctness ratio of the open-ended setting is $0.84$, which is sufficiently high. 
 We analyzed the background of those workers who submitted incorrect examples. Most of them use Facebook often (40\%, 40\%, and 20\% use Facebook every day, every week, and every month). 
The average self-evaluation of privacy scale of those workers who submitted incorrect examples is 2.8 (out of 5). This shows that the workers' background knowledge indeed impact the correctness of their ideas. 
 We also did the same measurement of the closed-ended setting. Out of thirty-six examples, thirteen examples are incorrect. The correctness ratio is $0.64$. Regarding the workers who made incorrect answers, most of them use Facebook often (42\%, 42\%, and 16\% use Facebook every day, every week, and every month). 71\% of them considered themselves adequately protected their own private information on Facebook. The average self-evaluation of their privacy scale is 3.4 (out of 5). Surprisingly the workers of the closed-ended experiments are more confident in their privacy knowledge than the open-ended experiments, while they produce more incorrect answers. This shows that the general crowd may over-estimate their understanding of privacy. 

\begin{figure}[!ht]
  \begin{center}
      \includegraphics[width=0.35\textwidth]{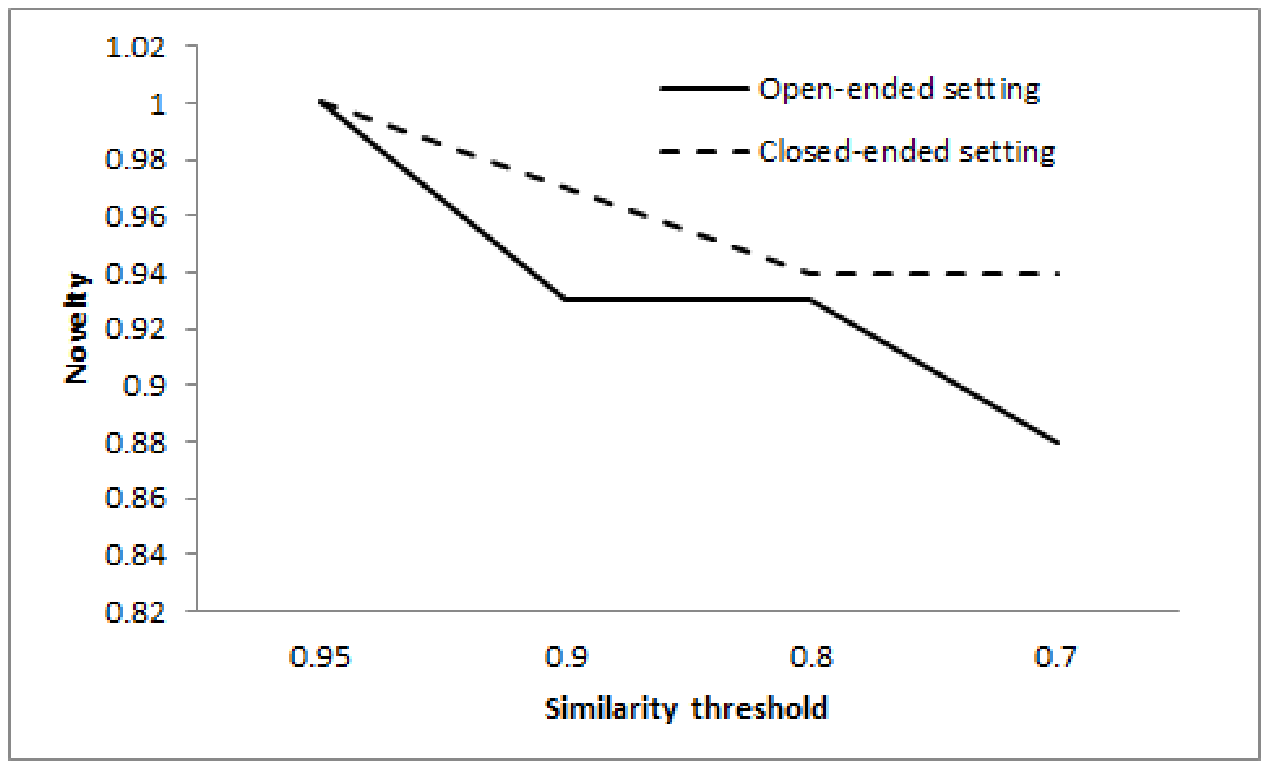}
    \caption{\small \label{fig:novelty} Novelty of Crowd-generated Ideas}
  \end{center}
\end{figure}

Second, we measure the novelty of the ideas. We choose different similarity thresholds (we use the same threshold value of $\delta$  and $\theta$). Figure \ref{fig:novelty} shows the results for both the open-ended and closed-ended settings. All reported results are computed as the average of novelty scores of all example postings. We have the following observations. First, the novelty reduces when the similarity threshold decreases. This is not surprising as higher similarity thresholds lead to fewer similar ideas, and thus higher novelty. Second, for both settings, the novelty is sufficiently high (at least 0.88). This shows that the crowd respected the creativity requirements in the HIT description, and generated high-novelty ideas. Third, the novelty of ideas generated in the closed-setting is higher than in the open-setting. This is not surprisingly as in the open-end setting, the workers incline to generate ideas on the popular privacy topics. These ideas have high probability to be similar. But in the closed-end setting, the picked privacy topics are those that the general public rarely pay attention to (as proven by the analysis of the open-end setting results in Section \ref{sc:integration}). Ideas generated on those topics are less likely to collude and more likely to be dissimilar. 
We also asked the privacy experts to review the ideas from the crowd regarding the novelty. The privacy experts agreed that some privacy examples with high novelty scores were indeed new and urging. 

Given both high correctness ratio and high novelty, we believe that in terms of game-based privacy learning, the non-expert crowd indeed can generate high-quality game design ideas from their experiences, as they may be able to generate ideas in new ways and may have access to solutions that the experts do not.  

\noindent{\bf Correlations.} To investigate whether there exists a relationship between the background of an individual and the quality of his generated ideas, we measure the Pearson correlation between the correctness \& novelty of ideas and the workers' privacy background \& experiences. 
We consider the workers' self-evaluation of expertise level as their privacy expertise, and their self-evaluation of whether having provided sufficient privacy protection on social networking sites as their privacy experience. 
The correlation results are shown in Table \ref{table:corr}. We have the following findings. First, the idea correctness is positively correlated to the workers' privacy expertise. This is not surprising as with more expertise the workers understand the problem better and thus produce more suitable answer. Second, the correlation between the idea correctness and the workers' online privacy experience is very weak. This is somewhat surprising but still can be explained: Getting more experiences may not be able to help the workers to understand privacy better. Sometimes the users repeat the same mistakes if they were not aware. 
Regarding the novelty of ideas, the correlation between the novelty and the expertise, as well as the correlation between the novelty and experience, were negative. This implies that indeed the workers' background does not help to generate novel ideas. We must admit that the correlation results may be biased, as the workers may over-estimate their expertise level. They may also be over-confident of their online privacy experience. 
\begin{table}
\begin{center}
\begin{tabular}{ |c|c|c|}\hline
            & Privacy expertise & Privacy experience \\\hline
Correctness &  0.42  & -0.05 \\\hline
Novelty     &  -0.22 & -0.32 \\\hline
\end{tabular}
\end{center}
\caption{\label{table:corr} Correlation between idea quality and Workers' background}
\end{table}

\subsection{Game Evaluation}

The game evaluation involves both crowd workers on Amazon Mechanic Turk (AMT) and three privacy experts. 
One hundred and twenty AMT workers participated into the evaluation task. Each worker is required to finish at least one game character, and answered all questions in the survey based on the character(s) that (s)he played with. The survey questions asked the workers to evaluate both the usability and the education success. We ensure that all characters were reviewed by the same number of workers by assigning four surveys (one for each character) to the same number of workers.

\begin{figure}[!ht]
  \begin{center}
	  \includegraphics[width=0.35\textwidth]{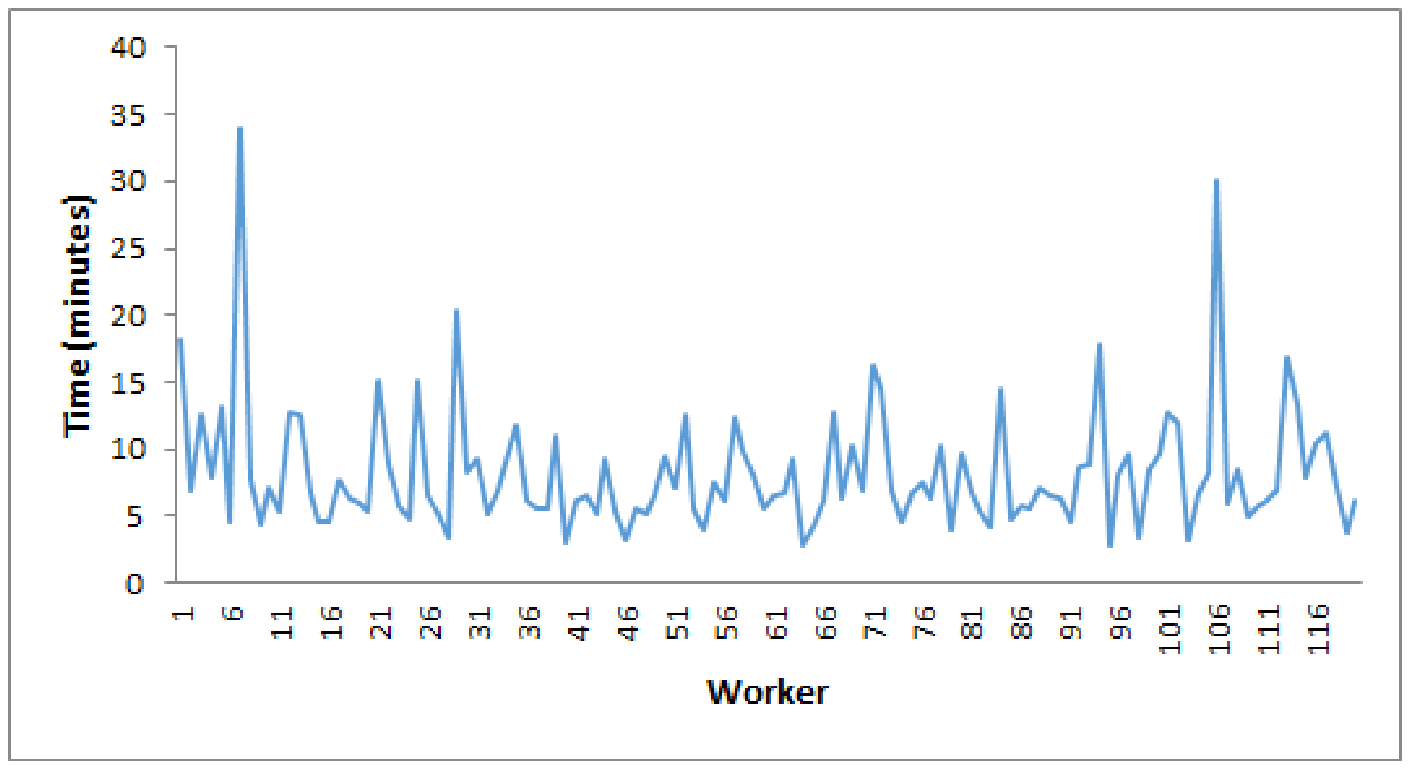}
		\caption{\small \label{fig:worktime} Time to Finish Survey}
  \end{center}
\end{figure}

\noindent{\bf Detecting cheating workers.}  We record the time that each participant took to finish the survey (including game playing). It turns out that the length of the working time is dramatically diverse. Figure \ref{fig:worktime} shows the distribution of the time length. The length of the working time varies from 2.78 minutes to 33.9 minutes. The average of working time length is 8.3 minutes. As our privacy experts spent two minutes at average to finish playing one game character, we expect that the survey should take at least four minutes to finish (including game playing). We consider those workers who finished the survey in less than four minutes as {\em cheating}. There are 10 (out of 120) cheating workers based on the time analysis. Therefore, we consider our collected evaluation results acceptable. 

\begin{figure*}[ht]
  \begin{center}
    \begin{tabular}{ccc}
      \includegraphics[width=0.3\textwidth]{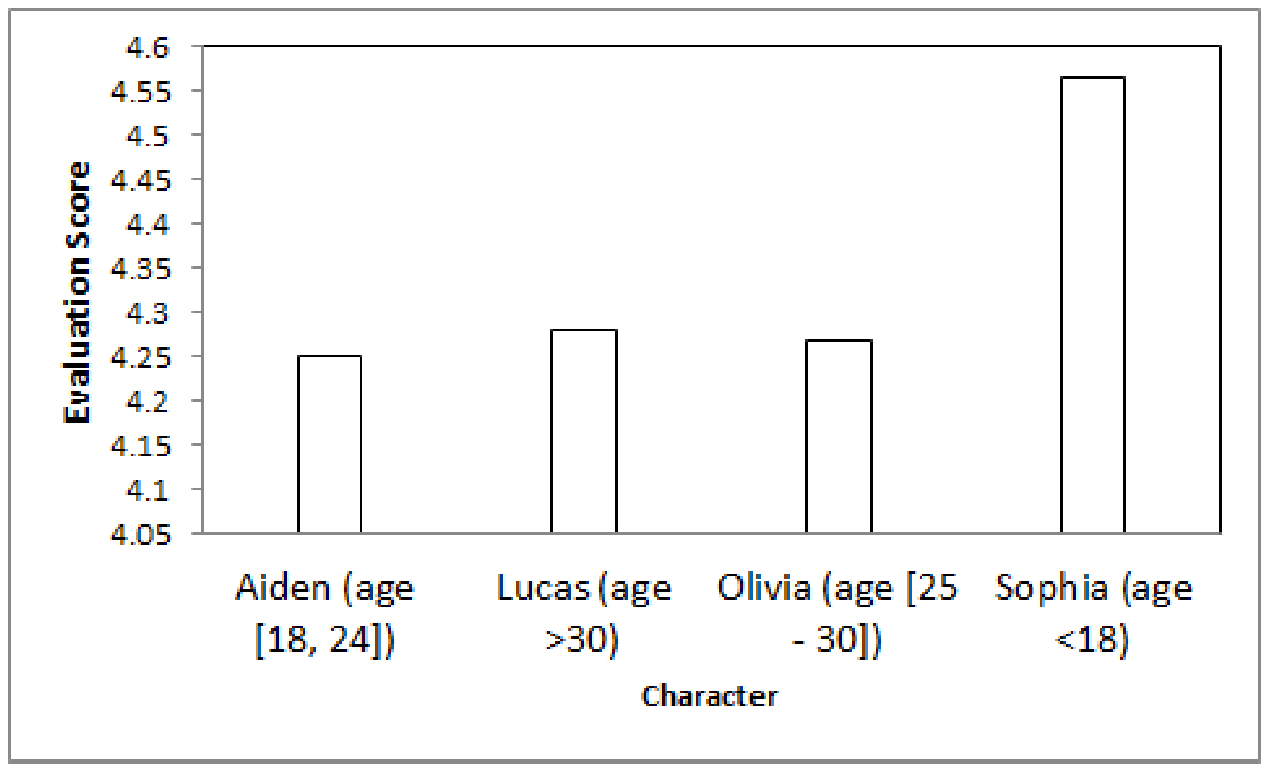}
      &
      \includegraphics[width=0.3\textwidth]{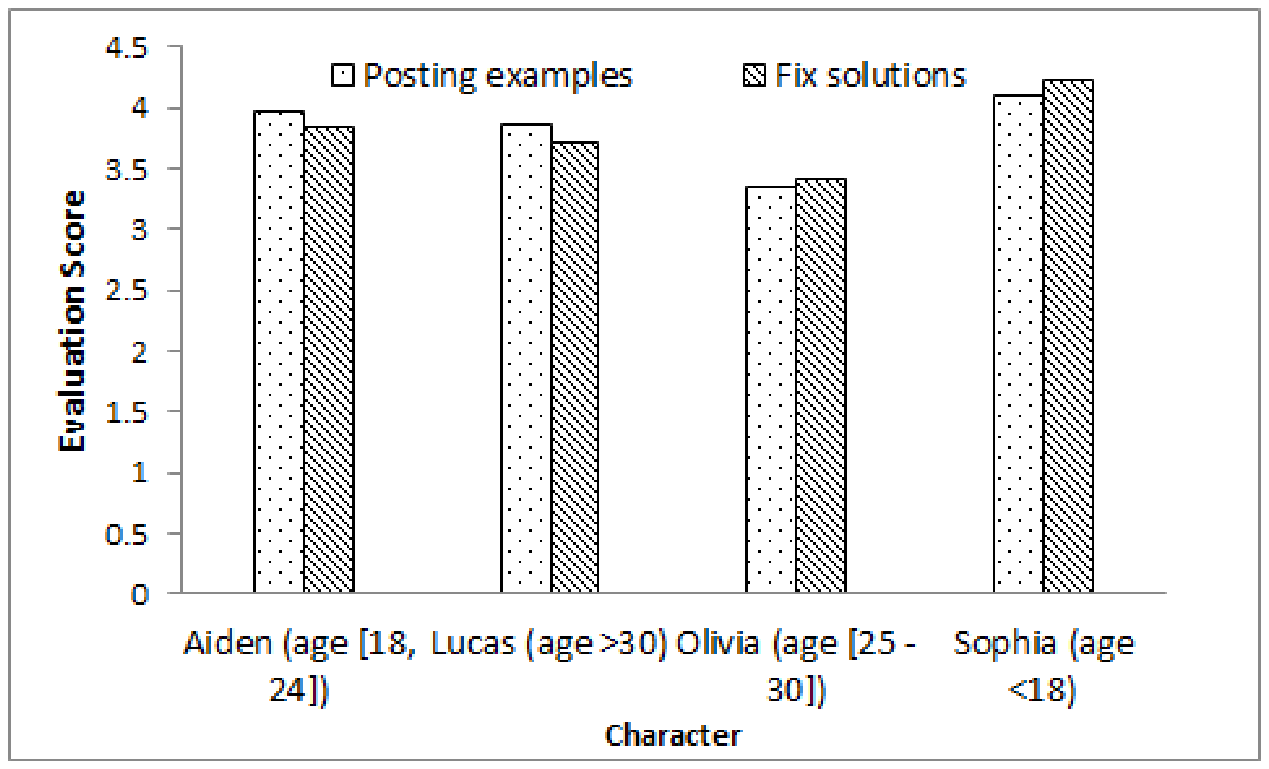}
			&
      \includegraphics[width=0.3\textwidth]{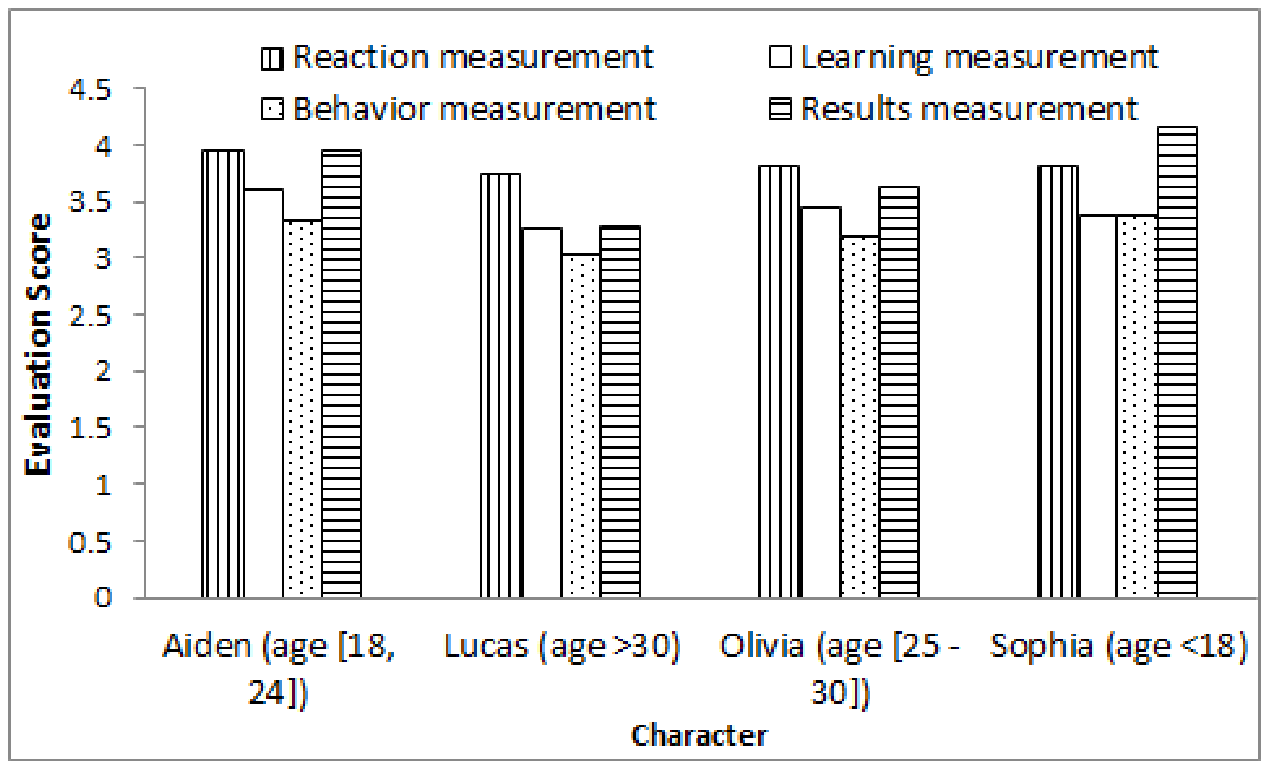}
      \\
      (a) Usability evaluation: game interface
      &
      (b) Usability evaluation: game content
			&
			(c) Evaluation of education success
    \end{tabular}
    \caption{\small \label{fig:eval} Game Evaluation}
  \end{center}
\end{figure*}

\noindent{\bf Tool usability.} The tool usability evaluation is performed in the format of survey. The survey includes the questions that ask users to score on both the game interface and the content (More details of the survey questions are in Section \ref{sc:eval}). The evaluation scores are in a scale of 1 to 5, with 1 being strongly disagree, and 5 being strongly agree.
We compute the average of evaluation scores per evaluation component per character. Figure \ref{fig:eval} (a) and (b) show the results of tool usability evaluation. Most of the components received the evaluation score no lower than 3.03. In particular, the crowd are satisfied with the game interface, with the evaluation score no lower than 4, for all four characters (Figure \ref{fig:eval} (a)). The crowd also enjoyed the examples and the suggested fix solutions (scores at least 3.36), for all four characters (Figure \ref{fig:eval} (b)). We compute the average score of the game interface and the content. The Sophia character receives the highest average score, while the Olivia character receives the lowest score. 

\noindent{\bf Education success.} The education success evaluation includes the grading of the four learning components, namely, reaction measurement, learning measurement, behavior measurement, and results measurement. The score scale is the same as those questions for tool usability evaluation. The evaluation results are shown in Figure \ref{fig:eval} (c). We compute the average score of the four components. It turned out that the average education score is in the range of $[3.23, 3.83]$. In other words, the crowd considers the game educational in general. From the scores of each individual component, we observe that the crowd agrees that the game works the best on the reaction measurement, but the worst on the behavior measurement. Regarding the education impact of different game characters, the Aiden character receives the highest average score for education, while the Olivia character receives the lowest score.

\noindent{\bf Crowd VS. experts.} We asked the privacy experts to play the game and finish the same survey as the AMT workers. We calculated the average scores of both usability and education success by the experts, and computed the difference between the average scores between the experts and the crowd. It turns out the experts' opinions are close to the crowd in terms of tool usability (average score difference $<0.3$). The experts also agree that the game is educational in general, but with scores 15\% lower than the crowd.  Surprisingly, the experts picked the character Lucas as the most educational, but Aiden the worst. We interviewed with the experts and found that the disagreement comes from some questions that the crowd labeled as inspiring but considered as trivial by the experts. This proves that indeed it is possible that the privacy experts may have different opinions from the general public regarding the effectiveness of the education game. 
\vspace{-0.1in}
\section{Related Work}
\label{sc:related}

Using the Internet and related technologies, crowdsourcing is able to access the brainpower of many people and achieve a wide range of outcomes, from the simple and mundane task of collecting street addresses of companies, to more sophisticated achievement, such as Wikipedia, innovation competitions, and helping solve cutting-edge scientific problems \cite{eiben2012increased,kittur2013future}. 
One type of tasks that are popular in the field of computer science is to combine human-generated results with computer algorithms to achieve synergistic outcomes \cite{kittur2013future}. For example, crowds' input has been used to improve automatic extraction of concepts in texts \cite{demartini2012zencrowd} and to improve search results from search engines \cite{bozzon2012answering}. This type of crowdsourcing builds the foundation for more advanced techniques for large scale human-computer collaboration.

Another popular type of crowdsourcing is to give creative tasks to a crowd. In this type, instead of having people generate some close-ended answers, people are explicitly told to generate novel outcomes. Yu and Nickerson \cite{yu2011cooks} conducted studies in which a crowd developed chair design sketches and successive crowds combined the designs of previous crowds, resembling genetic algorithm in computer science. 
The authors showed that crowd members tended to integrate novel and practical features of design, which help improve creativity \cite{yu2011feature}. In another study, crowd-based genetic algorithm was used to generate graphical advertisements of an online game. The results showed that having crowds modifying previous generation of ads generated better ads than having crowds combining previous ads \cite{ren2014increasing}. 

In addition to making design sketches, textual idea generation is also a popular creative task in crowdsourcing research. An important question is whether crowds are able to generate ideas of similar quality as professionals. Poetz and Schreirer \cite{poetz2012value} show that product ideas from an idea contest with customers have higher novelty and customer benefit than professionals' ideas, although the idea feasibility is somewhat lower. Some top ideas that the executives like are from customers, or crowds. Therefore, it is possible for crowds to generate useful ideas, especially if they are the end users of the products. Some specific technique in crowdsourcing idea generation show their effectiveness in improving idea creativity, such as deliberately finding source of analogies from other websites \cite{yu2014searching} and decomposing the initial creative task into  sub-problems \cite{luo2015improving}. 


In addition to idea generation, idea evaluation is also an important topic in crowdsourcing creative tasks. It is found that using multiple scales (e.g., novelty, value, feasibility) to measure idea quality is beneficial if ideas are not long \cite{riedl2013effect}. If ideas are long, a single scale measurement (only idea quality) may lead to more accurate evaluation. The study also finds that having 20 ratings for an idea generates accurate evaluation. In another study of crowd-generated ideas, it is found that prediction voting (predicting whether an idea can win the competition) is more appropriate when many poor ideas need to be filtered out, while Likert scale ratings are more appropriate when more refined distinctions need to be made for ideas that are of reasonable quality \cite{Sakamoto-amcis11}.

\section{Conclusion}
\label{sc:conclusion}

In this paper, we present our effort of leveraging crowdsourcing techniques for game-based privacy learning. We utilize the crowd for both idea collection for the game design and game evaluation. Our experiments show that the non-expert crowd can provide useful inputs for both game design and evaluations.

For the future work, we plan to improve $\crowdpet$ based on the received feedback from the {\em game evaluation} phase. We also plan to explore if under certain assumptions the ``wisdom of the crowd'' is able to outperform a smaller group of ``experts''.  Longer term, we want to investigate the potential of recruiting non-expert workers to collaborate on designing 
large-scale content-creation projects. 

\section{Acknowledgment}
This material is based upon work supported by the National Science Foundation under Grant \# SaTC-1464800. Any opinions, findings, and conclusions or recommendations expressed in this material are those of the author(s) and do not necessarily reflect the views of the National Science Foundation. 

\bibliographystyle{abbrv}
{
\bibliography{depetbib,bib}

\begin{thebibliography}{10}

\bibitem{turk}
Amazon merchanic turk.
\newblock https://www.mturk.com/mturk/welcome.

\bibitem{sheep}
The sheep market.
\newblock http://www.thesheepmarket.com/.

\bibitem{social}
Social networking fact sheet.
\newblock http://www.pewinternet.org/fact-sheets/social-networking-fact-sheet/.

\bibitem{Albrechtslund-08}
A.~Albrechtslund.
\newblock Online social networking as participatory surveillance.
\newblock {\em First Monday}, 13(3), 2008.

\bibitem{alonso2008crowdsourcing}
O.~Alonso, D.~E. Rose, and B.~Stewart.
\newblock Crowdsourcing for relevance evaluation.
\newblock In {\em ACM SigIR Forum}, volume~42, pages 9--15, 2008.

\bibitem{Sakamoto-amcis11}
J.~Bao, Y.~Sakamoto, and J.~V. Nickerson.
\newblock Evaluating design solutions using crowds.
\newblock In {\em AMCIS 2011 Proceedings - All Submissions. Paper 446}.
  Association for Information Systems, 2011.

\bibitem{blei2003latent}
D.~M. Blei, A.~Y. Ng, and M.~I. Jordan.
\newblock Latent dirichlet allocation.
\newblock {\em the Journal of machine Learning research}, 3:993--1022, 2003.

\bibitem{bozzon2012answering}
A.~Bozzon, M.~Brambilla, and S.~Ceri.
\newblock Answering search queries with crowdsearcher.
\newblock In {\em Proceedings of the 21st international conference on World
  Wide Web}, pages 1009--1018, 2012.

\bibitem{brabham2009moving}
D.~C. Brabham.
\newblock Moving the crowd at threadless: Motivations for participation in a
  crowdsourcing application.
\newblock 2009.

\bibitem{brooks2010design}
F.~P. Brooks~Jr.
\newblock {\em The design of design: Essays from a computer scientist}.
\newblock Pearson Education, 2010.

\bibitem{demartini2012zencrowd}
G.~Demartini, D.~E. Difallah, and P.~Cudr{\'e}-Mauroux.
\newblock Zencrowd: leveraging probabilistic reasoning and crowdsourcing
  techniques for large-scale entity linking.
\newblock In {\em Proceedings of the 21st international conference on World
  Wide Web}, pages 469--478, 2012.

\bibitem{eiben2012increased}
C.~B. Eiben, J.~B. Siegel, J.~B. Bale, S.~Cooper, F.~Khatib, B.~W. Shen,
  F.~Players, B.~L. Stoddard, Z.~Popovic, and D.~Baker.
\newblock Increased diels-alderase activity through backbone remodeling guided
  by foldit players.
\newblock {\em Nature biotechnology}, 30(2):190--192, 2012.

\bibitem{hargittai2010facebook}
E.~Hargittai and D.~Boyd.
\newblock Facebook privacy settings: Who cares?
\newblock {\em First Monday}, 15(8), 2010.

\bibitem{huangprivacy}
Y.~Huang, Y.~Wang, and Y.~Tang.
\newblock Privacy in emotion sharing on social media.
\newblock In {\em Proceedings of the Symposium on Usable Privacy and Security
  (SOUPS)}, 2014.

\bibitem{kazai2011crowdsourcing}
G.~Kazai, J.~Kamps, M.~Koolen, and N.~Milic-Frayling.
\newblock Crowdsourcing for book search evaluation: impact of hit design on
  comparative system ranking.
\newblock In {\em Proceedings of the 34th international ACM SIGIR conference on
  Research and development in Information Retrieval}, pages 205--214, 2011.

\bibitem{kirkpatrick-1979techniques}
D.~L. Kirkpatrick.
\newblock Techniques for evaluating training programs.
\newblock {\em Classic writings on instructional technology}, 1:231--241, 1979.

\bibitem{kittur2013future}
A.~Kittur, J.~V. Nickerson, M.~Bernstein, E.~Gerber, A.~Shaw, J.~Zimmerman,
  M.~Lease, and J.~Horton.
\newblock The future of crowd work.
\newblock In {\em Proceedings of the 2013 conference on Computer supported
  cooperative work}, pages 1301--1318, 2013.

\bibitem{kittur2011crowdforge}
A.~Kittur, B.~Smus, S.~Khamkar, and R.~E. Kraut.
\newblock Crowdforge: Crowdsourcing complex work.
\newblock In {\em Proceedings of the 24th annual ACM symposium on User
  interface software and technology}, pages 43--52, 2011.

\bibitem{koudas2006record}
N.~Koudas, S.~Sarawagi, and D.~Srivastava.
\newblock Record linkage: similarity measures and algorithms.
\newblock In {\em Proceedings of the 2006 ACM SIGMOD international conference
  on Management of data}, pages 802--803, 2006.

\bibitem{sung99}
S.~H. Lee.
\newblock Usability testing for developing effective interactive multimedia
  software: Concepts, dimensions, and procedures.
\newblock {\em Educational Technology \& Society}, (2), 1999.

\bibitem{leimeister2009leveraging}
J.~M. Leimeister, M.~Huber, U.~Bretschneider, and H.~Krcmar.
\newblock Leveraging crowdsourcing: activation-supporting components for
  it-based ideas competition.
\newblock {\em Journal of management information systems}, 26(1):197--224,
  2009.

\bibitem{Li-cogsci13}
H.~Li and Y.~Sakamoto.
\newblock The influence of collective opinion on true-false judgment and
  information-sharing decision.
\newblock In {\em Annual Meeting of the Cognitive Science Society}. Cognitive
  Science Society, 2013.

\bibitem{luo2015improving}
L.~Luo and O.~Toubia.
\newblock Improving online idea generation platforms and customizing the task
  structure based on consumers' domain specific knowledge.
\newblock {\em Journal of Marketing}, 2015.

\bibitem{online}
H.~McCracken.
\newblock Teaching with technology: Tools and strategies to improve student
  learning.
\newblock {\em Faculty Focus}, 2011.

\bibitem{papadimitriou1998latent}
C.~H. Papadimitriou, H.~Tamaki, P.~Raghavan, and S.~Vempala.
\newblock Latent semantic indexing: A probabilistic analysis.
\newblock In {\em Proceedings of the seventeenth ACM SIGACT-SIGMOD-SIGART
  symposium on Principles of database systems}, pages 159--168, 1998.

\bibitem{pass04}
F.~Pass and O.~Firssova.
\newblock {\em Usability Evaluation of Integrated E-Learning}.
\newblock New York: RoutledgeFalmer (Taylor \& Francis Group), 2004.

\bibitem{poetz2012value}
M.~K. Poetz and M.~Schreier.
\newblock The value of crowdsourcing: can users really compete with
  professionals in generating new product ideas?
\newblock {\em Journal of Product Innovation Management}, 29(2):245--256, 2012.

\bibitem{prensky2005computer}
M.~Prensky.
\newblock Computer games and learning: Digital game-based learning.
\newblock {\em Handbook of computer game studies}, 18:97--122, 2005.

\bibitem{ren2014increasing}
J.~Ren, J.~V. Nickerson, W.~Mason, Y.~Sakamoto, and B.~Graber.
\newblock Increasing the crowd's capacity to create: how alternative generation
  affects the diversity, relevance and effectiveness of generated ads.
\newblock {\em Decision Support Systems}, 65:28--39, 2014.

\bibitem{riedl2013effect}
C.~Riedl, I.~Blohm, J.~M. Leimeister, and H.~Krcmar.
\newblock The effect of rating scales on decision quality and user attitudes in
  online innovation communities.
\newblock {\em International Journal of Electronic Commerce}, 17(3):7--36,
  2013.

\bibitem{Sakamoto-icis11}
Y.~Sakamoto and J.~Bao.
\newblock Testing tournament selection in creative problem solving using
  crowds.
\newblock In {\em Proceedings of the International Conference on Information
  Systems (ICIS)}, 2011.

\bibitem{shiratuddin02}
N.~Shiratuddin and M.~Landoni.
\newblock Evaluation of content activities in children's educational software.
\newblock {\em Evaluation and Program Planning}, 25(2):175–--182, 2002.

\bibitem{sisarica2013emerging}
A.~Sisarica and N.~Maiden.
\newblock An emerging model of creative game-based learning.
\newblock In {\em Proceedings of the International Conference of Serious Games
  Development and Applications (SGDA)}, pages 254--259. 2013.

\bibitem{yu2014searching}
L.~Yu, A.~Kittur, and R.~E. Kraut.
\newblock Searching for analogical ideas with crowds.
\newblock In {\em Proceedings of the 32nd annual ACM conference on Human
  factors in computing systems}, pages 1225--1234, 2014.

\bibitem{yu2011cooks}
L.~Yu and J.~V. Nickerson.
\newblock Cooks or cobblers?: crowd creativity through combination.
\newblock In {\em Proceedings of the SIGCHI conference on human factors in
  computing systems}, pages 1393--1402, 2011.

\bibitem{yu2011feature}
L.~Yu and Y.~Sakamoto.
\newblock Feature selection in crowd creativity.
\newblock In {\em Foundations of Augmented Cognition. Directing the Future of
  Adaptive Systems}, pages 383--392. 2011.

\bibitem{zeichner2012crowdsourcing}
N.~Zeichner, J.~Berant, and I.~Dagan.
\newblock Crowdsourcing inference-rule evaluation.
\newblock In {\em Proceedings of the 50th Annual Meeting of the Association for
  Computational Linguistics: Short Papers-Volume 2}, pages 156--160, 2012.

\end{thebibliography}
}

\end{document}